\documentclass[aps,pra,reprint,groupedaddress,longbibliography,nofootinbib]{revtex4-1}
\usepackage{graphicx,amsmath,amssymb,braket,siunitx,mathtools}
\usepackage[unicode]{hyperref}
\usepackage[dvipsnames]{xcolor}
\hypersetup{
    colorlinks=true,
    citecolor=blue,
    linkcolor=blue,
    urlcolor=blue
}
\DeclareSIUnit\gauss{G}
\DeclareSIUnit\sig{\mbox{$\sigma$}}

\DeclarePairedDelimiterX\cusket[1]{\vert}{\rangle}{#1}
\DeclarePairedDelimiterX\cusbra[1]{\langle}{\vert}{#1}
\DeclarePairedDelimiterX\cusbraket[2]{\langle}{\rangle}{#1 \delimsize\vert #2}
\newcommand{\sbraket}[4]{%
  {\hspace{-0.1666em}_{~{_#2}\hspace{-0.25em}\vphantom{\smash[b]{|}}}\cusbraket{#1}{#3}_{\vphantom{\smash[b]{|}}\!{_#4}}}
}
\newcommand{\sket}[2]{%
  {\cusket{#1}_{\vphantom{\smash[b]{|}}\!{_#2}}}
}
\newcommand{\sbra}[2]{%
  {\hspace{-0.1666em}_{~{_#2}\hspace{-0.25em}\vphantom{\smash[b]{|}}}\cusbra{#1}}
}

\newcommand{\myol}[2][3]{{}\mkern#1mu\overline{\mkern-#1mu#2}}

\begin{document}

\title{Sideband cooling of molecules in optical traps}
\author{L. Caldwell}
\author{M. R. Tarbutt}
\email[]{m.tarbutt@imperial.ac.uk}
\affiliation{Centre for Cold Matter, Blackett Laboratory, Imperial College London, Prince Consort Road, London SW7 2AZ UK
}

\begin{abstract}
Sideband cooling is a popular method for cooling atoms to the ground state of an optical trap. Applying the same method to molecules requires a number of challenges to be overcome. Strong tensor Stark shifts in molecules cause the optical trapping potential, and corresponding trap frequency, to depend strongly on rotational, hyperfine and Zeeman state. Consequently, transition frequencies depend on the motional quantum number and there are additional heating mechanisms, either of which can be fatal for an effective sideband cooling scheme. We develop the theory of sideband cooling in state-dependent potentials, and derive an expression for the heating due to photon scattering.  We calculate the ac Stark shifts of molecular states in the presence of a magnetic field, and for any polarization. We show that the complexity of sideband cooling can be greatly reduced by applying a large magnetic field to eliminate electron- and nuclear-spin degrees of freedom from the problem. We consider how large the magnetic field needs to be, show that heating can be managed sufficiently well, and present a simple recipe for cooling to the ground state of motion.
\end{abstract}


\maketitle

\section{Introduction}
\label{sec:intro}

In recent years there has been rapid progress in the development of techniques for producing and manipulating ultracold molecules \cite{Ni2008, Danzl2008, Barry2014, Prehn2016, Truppe2017, Park2017, Moses2017, Collopy2018, Williams2018, Cheuk2018, Caldwell2019, DeMarco2019}. Arrays of molecules interacting via the dipole-dipole interaction can be used as a platform to study many-body quantum physics \cite{Micheli2006,Micheli2007, Capogrosso-Sansone2010, Sowinski2012} or to implement two-qubit quantum gates \cite{Ni2018}. Small arrays can be made using tweezer traps, and larger arrays using optical lattices. Molecules produced by association of ultracold atoms have been loaded into lattices at high enough filling factors to begin studying many-body effects \cite{Yan2013}. Molecules have also been formed by associating pairs of atoms in tweezer traps \cite{Liu2018}.

Very recently, laser-cooled molecules were captured in tweezer traps for the first time \cite{Anderegg2019}. To exploit the potential of these low-entropy arrays it is necessary to initialize each molecule in a single quantum state. An important current challenge is how to cool these molecules to the ground state of motion in tweezer traps or lattices. This is frequently done for alkali atoms using Raman sideband cooling \cite{Kaufman2012, Yu2018}, and these methods are now being extended to alkaline-earth atoms \cite{Norcia2018, Cooper2018}. Application of the same techniques to molecules is difficult because (i) the complex structure of the molecule tends to complicate all laser cooling methods and (ii) molecules have large tensor Stark shifts, resulting in state-dependent trapping potentials that, in most circumstances, make sideband cooling impossible.

\begin{figure}
    \centering
    \includegraphics{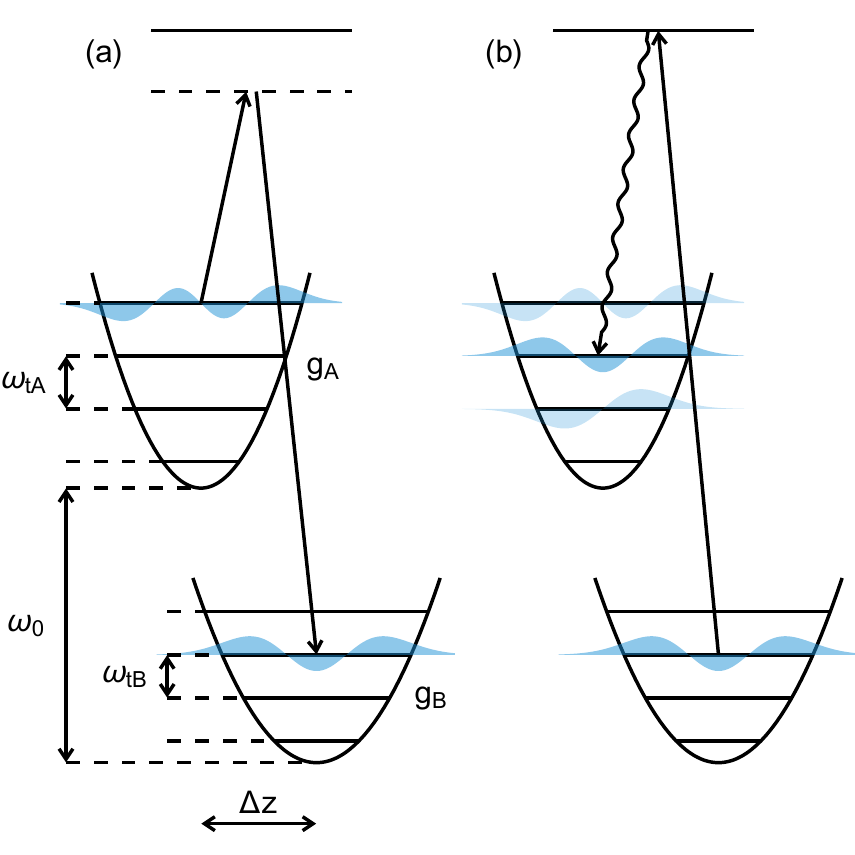}
    \caption{Schematic of the two steps of Raman sideband cooling. (a) Step 1: two-photon Raman transition with detuning set to change internal state from $g_A$ to $g_B$ whilst removing one quantum of motional energy. This step is forbidden when the molecule reaches the motional ground state. (b) Step 2: optical pumping via an electronic excited state returns the molecule to its original internal state. For cooling to work efficiently, the optical pumping should preserve the motional quantum number.}
    \label{fig:rsc-diagram}
\end{figure}

Raman sideband cooling consists of a repeated two-step process illustrated in Fig.~\ref{fig:rsc-diagram}. The first step drives a stimulated two-photon transition between a pair of internal states, reducing the motional quantum number $n$. In order to selectively drive the red sideband of the transition, the linewidth must be narrow compared to the energy spacing of the motional levels of the trapped atom; typical trap frequencies are of order \SI{100}{\kilo\hertz}. The second step provides the dissipation necessary for cooling by optically pumping the atom back to its original internal state.

There are two key requirements for effective cooling. Both are challenging for molecules. First, in order to cool from a thermal state, the frequency required to drive the stimulated transition must be independent of the motional state of the molecule. In alkali atoms---which typically have very small tensor Stark shifts\footnote{It is sometimes stated that the tensor Stark shift is zero for ground-state alkali atoms. This is only true in the absence of hyperfine structure.} \cite{LeKien2013}---it is straightforward to find pairs of internal states for which the trapping potentials are nearly identical. Provided the potential is sufficiently harmonic, the transition frequency is then independent of the initial motional state. Because molecules have large tensor Stark shifts, the trap frequency depends strongly on internal state, so the two-photon transition frequency depends on the motional state.

The second requirement is that the optical pumping step must have a high probability of preserving the motional quantum number. When the potentials are identical, different motional states of the two potentials are orthogonal and the probability of changing $n$ depends only on the Lamb-Dicke parameter, the square root of the ratio of the photon recoil energy to the level spacing of the traps. Provided this parameter is small, transitions that change the motional quantum number are strongly suppressed. However, for state-dependent potentials, the different motional states of the two traps are no longer orthogonal and the heating involved in scattering a photon has additional contributions associated with projection of the molecule from one potential to the other. Moreover, the additional complexity of a molecule compared to an atom means that optical pumping to a desired state often requires more photons to be scattered, each contributing to the heating.

A potential advantage of state-dependent potentials is that they could enable projection cooling schemes \cite{Li2012, Weiss1997}. For example, one could drive a rotational transition resonant only with molecules in the motional ground state followed by state-dependent detection to determine whether or not the molecule has made the transition. If it has, the molecule has been projected into the motional ground state, if it has not we can reapply cooling light to scramble the motional state and try again. Whilst such schemes may be useful for a single molecule, they scale poorly and so are not suitable for arrays. An active cooling method is required.

This paper is organized as follows. In Sec.~\ref{sec:sideband-cooling-theory} we outline the theory of sideband cooling in state-dependent harmonic potentials and establish a quantitative set of criteria for efficient cooling to take place. In Sec.~\ref{sec:stark-op} we outline the effective Stark shift operator for the interaction of molecules with the trapping light. The results are applied in Sec.~\ref{sec:simple-molecule} to determine the energy levels of a simplified molecule in an idealized tweezer trap and consider how they can be engineered to meet the requirements for sideband cooling. In Sec.~\ref{sec:real-molecule} we consider complications that arise when we include the complex structure of real molecules---we use CaF as a case study, since it has already been loaded into a tweezer trap~\cite{Anderegg2019}. In Sec.~\ref{sec:real-tweezer} we discuss potential complications arising from the light field produced by a real tweezer trap. Finally, in Sec.~\ref{sec:conclusions}, we propose a complete recipe for Raman sideband cooling of laser-cooled molecules. The coordinate system and set of angles that we use throughout the paper are illustrated in Fig.~\ref{fig:coords}.

\section{Theory of sideband cooling in state-dependent potentials}
\label{sec:sideband-cooling-theory}

In this section, we develop the general theory of sideband cooling in state-dependent potentials in order to derive expressions for the frequency and amplitude of the transitions, and the heating arising from the optical pumping step. The Hamiltonian $H_0$ describing a molecule with a pair of ground states $\ket{g_A}$ and $\ket{g_B}$ in a state-dependent harmonic trap is
\begin{equation}
\begin{split}
    H_{\operatorname{2-level}} = &H_{\mathrm{t}A}\ket{g_A}\bra{g_A} + H_{\mathrm{t}B}\ket{g_B}\bra{g_B} \\
    &\quad+ \frac{\hbar \omega_0}{2}(\ket{g_A}\bra{g_A} - \ket{g_B}\bra{g_B}).
\end{split}
\end{equation}Here $H_{\mathrm{t}i}$ is the harmonic oscillator Hamiltonian associated with the external motion of a molecule in internal state $\ket{g_i}$, $i\in\{A,B\}$. The trap frequencies are $\omega_{\mathrm{t}i}$ and the motional eigenstates are $\sket{q}{i}$ such that $H_{\mathrm{t}i}\sket{q}{i} = \hbar \omega_{\mathrm{t}i} (q+1/2)\sket{q}{i}$. $\hbar \omega_0$ is the energy difference between the minima of the two trapping potentials, shown in Fig.~\ref{fig:rsc-diagram}.  

\subsection{Raman step}

In the first step of Raman sideband cooling, shown in Fig.~\ref{fig:rsc-diagram}(a), the two photon detuning required to coherently transfer the molecule from $\ket{g_A}\sket{n}{A}$ to $\ket{g_B}\sket{n-1}{B}$ is
\begin{equation}
\begin{split}
    \Delta_{\mathrm{coh}} &= \omega_0 + (n+\frac{1}{2})\omega_{\mathrm{t}A} - (n-1+\frac{1}{2})\omega_{\mathrm{t}B},\\
    &= \omega_0 + n(\omega_{\mathrm{t}A}-\omega_{\mathrm{t}B})+\frac{1}{2}(\omega_{\mathrm{t}A}+\omega_{\mathrm{t}B}),
\end{split}
\end{equation}
where we have assumed that the energy of $\ket{g_A}$ is above that of $\ket{g_B}$. We see that when the trap frequencies are different for the two internal states, $\Delta_\mathrm{coh}$ depends on the motional quantum number $n$.

The matrix element for a transition between the two internal states via interaction with a light field is proportional to $\sbra{m}{B}e^{i \Delta k z}\sket{n}{A}$ where $\hbar \Delta k$ is the momentum kick imparted to the molecule from absorption of the photon. For two-photon transitions, $\hbar \Delta k$ is the difference in momenta between the absorbed and emitted photons. We can re-express $\Delta k z$ as
\begin{equation}
    \begin{split}
        \Delta k z &= \Delta k \sqrt{\frac{\hbar}{2 M \omega_{\mathrm{t}A}}}(a_A^\dagger + a_A),\\
        &= \eta_A (a_A^\dagger + a_A),
    \end{split}
\end{equation}
where $a_i^\dagger$, $a_i$ are the harmonic oscillator raising and lowering operators associated with $H_{\mathrm{t}i}$ and $M$ is the mass of the molecule. We have also introduced the Lamb-Dicke parameter $\eta_i = \sqrt{E_{\rm rec} /\hbar \omega_{\mathrm{t}i}}$, the square root of the ratio of the recoil energy along the trap axis $E_{\rm rec}  = \hbar^2 \Delta k^2/2 M$, to the energy spacing of the motional states of the trap. When this ratio is small, commonly referred to as the Lamb-Dicke regime, we can expand the matrix element in powers of $\eta_A$,
\begin{equation}
\begin{split}
    \sbra{m}{B}e^{i \Delta k z}\sket{n}{A} =& \sum_{\ell=0}^\infty \frac{(i\eta_A)^\ell}{\ell!}\sbra{m}{B} (a_A^\dagger + a_A)^\ell\sket{n}{A}\\
    \simeq& \sbraket{m}{B}{n}{A} + i\eta_A(\sqrt{n}\sbraket{m}{B}{n-1}{A}\\
    &+\sqrt{n+1}\sbraket{m}{B}{n+1}{A})+\mathcal{O}(\eta_A^2).
\end{split}
\end{equation}
When the potentials associated with the two states are identical we have $\sbraket{m}{B}{n}{A}=\delta_{m,n}$ and the transition strength, $\left|\sbra{m}{B}e^{i \Delta k z}\sket{n}{A}\right|^2$, is proportional to $\eta_A^{2|m-n|}$. Under these conditions, transitions that change the motional quantum number are strongly suppressed. By using a two-photon Raman transition as shown in Fig.~\ref{fig:rsc-diagram}(a), $\Delta k$, and therefore $\eta_i$, can be varied by changing the relative directions of the two photons. Counter-propagating optical photons give sufficiently large $\Delta k$ to allow higher order sidebands to be addressed. In general the two harmonic oscillator potentials will have different trap frequencies and different equilibrium positions; an explicit expression for the overlap integral in this case can be found in \cite{Waldenstrom1982}.

\subsection{Optical pumping step}

\begin{figure}
    \centering
    \includegraphics{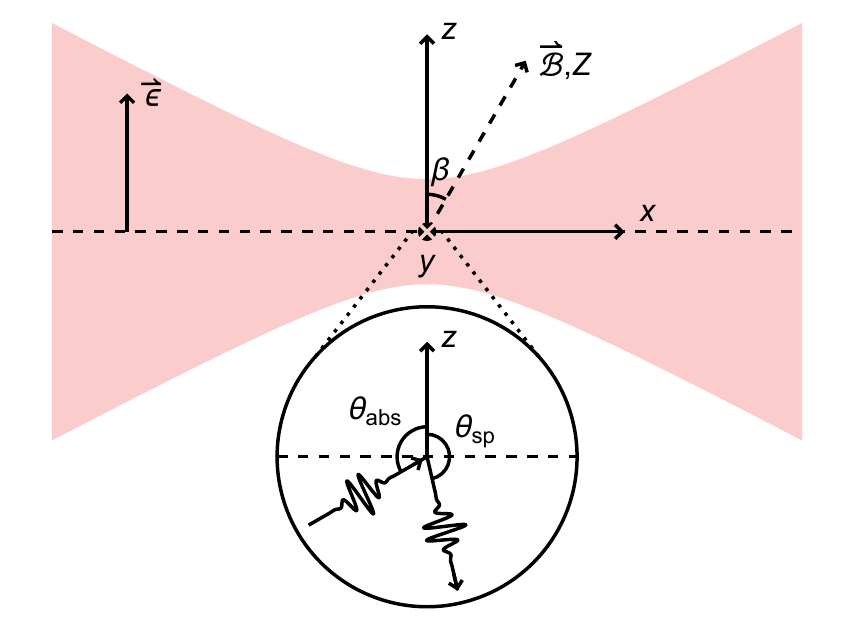}
    \caption{Schematic of the coordinate system used in this paper. The tweezer light propagates along $x$ and is (except where otherwise specified) linearly polarized along $z$. The $y$ axis is into the page. The quantization axis of the molecule $Z$, taken along the $\mathcal{B}$ field in Sec.~\ref{sec:real-molecule} onwards, makes an angle $\beta$ to the $z$ axis. Inset: For the optical pumping step we consider the angles $\theta_{\rm abs}$ and $\theta_{\rm sp}$ that the incoming and outgoing photons make with one of the three trap axes, chosen to be $z$ in our illustration.}
    \label{fig:coords}
\end{figure}

We now turn to the optical pumping step of the cooling cycle, which involves spontaneous emission. Consider the photon scattering event illustrated in Fig.~\ref{fig:rsc-diagram}(b). A particle in $\ket{g_B}\sket{n}{B}$ absorbs a photon from the laser and then decays to $\ket{g_A}$ as it spontaneously emits a photon. The angles of the incoming and outgoing photons relative to the trap axis are labeled $\theta_{\rm abs}$ and $\theta_{\rm sp}$, and are as shown in the inset of Fig.~\ref{fig:coords}. The probability of ending up in state $\sket{m}{A}$ is $\left|\sbra{m}{A}e^{i \Delta k z}\sket{n}{B}\right|^2$ where $\Delta k$ depends implicitly on $\theta_{\rm abs}$ and $\theta_{\rm sp}$.\footnote{The operator $e^{i \Delta k z}$ is unitary and so $\left|\sbra{m}{A}e^{i \Delta k z}\sket{n}{B}\right|^2$ gives a normalized probability: $\sum_m\left|\sbra{m}{A}e^{i \Delta k z}\sket{n}{B}\right|^2 = 1$.} For given directions of the incoming and outgoing photons, the mean change in motional quantum number is $\sum_m (m-n)\left|\sbra{m}{A}e^{i \Delta k z}\sket{n}{B}\right|^2$. Averaging over all possible directions of spontaneous emission gives
\begin{widetext}
\begin{equation}
    \begin{split}
        \overline{\Delta n^{\rm sc}_{B,A}} &= \frac{1}{2}\int_0^\pi\mathrm{d}\theta_{\rm sp}\mathcal{Y}(\theta_{\rm sp})\sin{\theta_{\rm sp}}\sum_m (m-n)\left|\sbra{m}{A}e^{i \Delta k z}\sket{n}{B}\right|^2\\
        &= \frac{1}{2}\int_0^\pi\mathrm{d}\theta_{\rm sp}\mathcal{Y}(\theta_{\rm sp})\sin{\theta_{\rm sp}}\sum_m \sbra{m}{A}\frac{1}{\hbar\omega_{\mathrm{t}A}}H_{\mathrm{t}A} e^{i \Delta k z} - \frac{1}{\hbar\omega_{\mathrm{t}B}}e^{i \Delta k z}H_{\mathrm{t}B} \sket{n}{B}\sbra{n}{B} e^{-i \Delta k z}\sket{m}{A},
        \end{split}
\end{equation}
where $\mathcal{Y}(\theta_{\rm sp})$ is the probability density for the photon to be emitted at angle $\theta_{\rm sp}$ to the trap axis.\footnote{If $\mathcal{G}(\theta)$ is the angular distribution of photon emission relative to the quantization axis, then $\mathcal{Y}(\theta_{\rm sp})=\int_0^{2\pi}\mathrm{d}\phi_{\rm sp}\left|\frac{\partial(\theta, \phi)}{\partial(\theta_{\rm sp}, \phi_{\rm sp})}\right|\mathcal{G}(\theta(\theta_{\rm sp}, \phi_{\rm sp}))$, with $\left|\frac{\partial(\theta, \phi)}{\partial(\theta_{\rm sp}, \phi_{\rm sp})}\right|$ the Jacobian determinant for the coordinate transformation which rotates the quantization axis on to the trap axis.} Using the completeness relation with respect to the set of $\sket{m}{A}$ we have
\begin{equation}
    \begin{split}
        \overline{\Delta n^{\rm sc}_{B,A}}&=\frac{1}{2}\int_0^\pi\mathrm{d}\theta_{\rm sp}\mathcal{Y}(\theta_{\rm sp})\sbra{n}{B}\frac{1}{\hbar\omega_{\mathrm{t}A}}e^{-i \Delta k z}H_{\mathrm{t}A} e^{i \Delta k z} - \frac{1}{\hbar\omega_{\mathrm{t}B}}H_{\mathrm{t}B} \sket{n}{B}\\
        &=\frac{1}{2}\int_0^\pi\mathrm{d}\theta_{\rm sp}\mathcal{Y}(\theta_{\rm sp})\left[\frac{1}{\hbar\omega_{\mathrm{t}A}}\sbra{n}{B}e^{-i \Delta k z}H_{\mathrm{t}A} e^{i \Delta k z}\sket{n}{B} - (n+\frac{1}{2})\right].
    \end{split}
\end{equation}
\end{widetext}
The remaining matrix element can be expanded as
\begin{equation}
    \begin{split}
        &\sbra{n}{B}e^{-i \Delta k z}H_{\mathrm{t}A} e^{i \Delta k z}\sket{n}{B}\\
        &= \sbra{n}{B}\frac{(p+\hbar \Delta k)^2}{2M}+\frac{1}{2}M \omega_{\mathrm{t}A}^2 (z-\Delta z)^2 \sket{n}{B}\\
        &=\frac{\hbar\omega_{\mathrm{t}B}}{2} (n+\frac{1}{2}) + E_{\rm rec} + \frac{\hbar \omega_{\mathrm{t}A}^2}{2 \omega_{\mathrm{t}B}}(n+\frac{1}{2}) + \frac{1}{2}M \omega_{\mathrm{t}A}^2 \Delta z^2,\label{eq:matrix-el}
    \end{split}
\end{equation}
where $\Delta z$ is the displacement between the minima of the two potentials. In the first step we have used $e^{-i \Delta k z}p e^{i \Delta k z}=p+\hbar\Delta k$ and in the last step $\sbra{n}{B}p^2\sket{n}{B}=\hbar M \omega_{\mathrm{t}B} (n+1/2)$, $\sbra{n}{B}z^2\sket{n}{B}=(\hbar/M \omega_{\mathrm{t}B})(n+1/2)$ and $\sbra{n}{B}p\sket{n}{B}=\sbra{n}{B}z\sket{n}{B}=0$. 

The second term in the last line of Eq.~(\ref{eq:matrix-el}), the recoil energy associated with the process, is the only part which depends on the directions of the absorbed and emitted photons. We define the average of this recoil energy over spontaneous emission directions,
\begin{equation}
\begin{split}
    \myol{E}_{\rm rec} =&\frac{1}{2}\frac{\hbar^2 k^2}{2 M} \int_0^{\pi}\left(\cos\theta_{\rm abs}-\cos\theta_{\rm sp}\right)^2 {\cal Y}(\theta_{\rm sp}) \sin\theta_{\rm sp}\mathrm{d}\theta_{\rm sp}\\
    =& \frac{\hbar^2 k^2}{2 M}(\cos^2\theta_{\rm abs} + \Upsilon),\label{eq:av-recoil-energy}
\end{split}
\end{equation}
where $\hbar k$ is the single photon momentum, and 
\begin{equation}
   \Upsilon = \frac{1}{2}\int_0^{\pi}\cos^2\theta_{\rm sp}{\cal Y}(\theta_{\rm sp}) \sin\theta_{\rm sp}\mathrm{d}\theta_{\rm sp}
\end{equation}
is a geometric factor that depends only on the polarization of the outgoing photon and the angle between the trap and quantization axes. Its value lies in the range $1/5\leq\Upsilon\leq2/5$. In the second step of Eq.~(\ref{eq:av-recoil-energy}), we have used the fact that $\mathcal{Y}(\theta_{\rm sp})$ is symmetric about $\theta_{\rm sp} = \pi/2$ so that the integral over the term linear in $\cos \theta_{\rm sp}$ is zero. Finally we can write  
\begin{equation}
    \begin{split}
        \overline{\Delta n^{\rm sc}_{B,A}} &= \Theta_{\mathrm{rec}} + \Theta_{\mathrm{disp}} + \Theta_{\mathrm{curv}},
    \end{split}
    \label{Eq:scatter-heating}
\end{equation}
where
\begin{subequations}
\begin{align}
    &\Theta_{\mathrm{rec}} = \frac{\myol{E}_{\rm rec} }{\hbar \omega_{\mathrm{t}A}} ,\label{eq:theta-recoil}\\
    &\Theta_{\mathrm{disp}} = \frac{1}{2}M \omega_{\mathrm{t}A}^2 \Delta z^2/(\hbar \omega_{\mathrm{t}A})\label{eq:theta-disp},\\
    &\Theta_{\mathrm{curv}} = \frac{1}{2}\left(\frac{\omega_{\mathrm{t}A}}{\omega_{\mathrm{t}B}} + \frac{\omega_{\mathrm{t}B}}{\omega_{\mathrm{t}A}}-2\right)(n+\frac{1}{2})\label{eq:theta-curv}.
\end{align}
\end{subequations}
The heating induced by the photon recoil, $\Theta_{\mathrm{rec}}$, is independent of $n$ and equivalent to the heating in free space or in state-independent potentials. The expression for $\myol{E}_{\rm rec}$ in Eq.~(\ref{eq:av-recoil-energy}) shows that this contribution to the heating can be split into a part due to the momentum of the absorbed photon and a part due to that of the spontaneously emitted photon. The distribution of the former among the three trap axes can be controlled by choosing the direction of the optical pumping beam. We note that the sum of $\myol{E}_{\rm rec}$ evaluated for any three perpendicular axes is $2\frac{\hbar^2 k^2}{2 M}$. The second contribution, $\Theta_{\mathrm{disp}}$, is the additional heating associated with the displacement between the two potentials. The quantity $\hbar\omega_{\mathrm{t}A}\Theta_{\mathrm{disp}}$ is the gain in potential energy from moving the wavepacket a distance $\Delta z$ up the side of the trap. Finally, $\Theta_{\mathrm{curv}}$ is the heating resulting from the difference in curvature of the two trap potentials. This part depends linearly on $n$ and is independent of the direction of the transition, $\ket{g_A}\leftrightarrow\ket{g_B}$.   

In general several photons will be scattered in the optical pumping step. Each scattering event begins with the molecule in some state $i$ and ends in some state $j$ with the associated mean change in motional quantum number $\overline{\Delta n^{\mathrm{sc}}_{i,j}}$. We define the mean change in $n$ for the complete process of optical pumping to the desired (dark) state, $\overline{\Delta n^{\mathrm{op}}}$, which is the sum of $\overline{\Delta n^{\mathrm{sc}}_{i,j}}$ for each step of the process. We will calculate $\overline{\Delta n^{\mathrm{op}}}$ for a realistic case in Sec.~\ref{sec:simple-molecule}. For efficient cooling, the number of motional quanta removed during the coherent step $\Delta n_\mathrm{coh}$, must be greater than $\overline{\Delta n^{\mathrm{op}}}$, remembering that the heating during optical pumping occurs for each axis regardless of which is being cooled during the coherent step. Whilst it is possible to use higher-order sidebands during the coherent step to satisfy this condition, cooling to the motional ground state requires $\Delta n_\mathrm{coh}=1$ since driving higher-order sidebands leaves population in other motional states.

\section{Stark shift}
\label{sec:stark-op}

To derive the potential for a molecule in a tweezer trap or an optical lattice, we need to understand its response to the trapping light. The interaction of a molecule with light is described by a term in the Hamiltonian $-\vec{d}\cdot \vec{E}$, where $\vec{d}$ is the dipole moment operator and $\vec{E}= \frac{1}{2} {\cal E}_0 (\hat{\epsilon} e^{-i \omega_{\rm L} t} + \hat{\epsilon}^{*} e^{i \omega_{\rm L} t})$ is the electric field of the light. Here, ${\cal E}_0$ is the electric field amplitude, $\omega_{\rm L}$ is the angular frequency of the light, and $\hat{\epsilon}$ is a unit polarization vector. We divide the complete Hamiltonian into a zeroth-order part, $H_0$, that describes the energy level structure of the molecule down to the rotational structure, and a part $H_1$ that describes level shifts smaller than the rotational splitting. $H_1$ includes the spin-rotation interaction, the hyperfine interaction, and the Zeeman and Stark interactions. We are interested in the small degenerate subspace of $H_0$ corresponding to a single rotational state. The effective Stark Hamiltonian that operates within this subspace is developed in the Appendix. It is
\begin{equation}
H_{\rm S}=-\frac{\mathcal{E}_0^2}{4}\sum _{K=0}^2 \sum _{P=-K}^K  (-1)^P {\mathcal A}^{K}_P \mathcal{P}_{-P}^K.
\label{Eq:HStark}
\end{equation}

Each of the three terms in the sum over $K$ in Eq.~(\ref{Eq:HStark}) is the scalar product of two rank-$K$ spherical tensors. The first, ${\mathcal A}^{K}$, is an operator related to the frequency-dependent polarizability of the molecule, and is given in terms of $\vec{d}$ and $\omega_{\rm L}$ by Eq.~(\ref{EqApp:AKP}). The second, $\mathcal{P}^K$, relates to the polarization of the light, and is given in terms of $\hat{\epsilon}$ by Eq.~(\ref{EqApp:PKP}). The matrix elements of ${\mathcal A}^{K}_{P}$ for $^{1}\Sigma$ and $^{2}\Sigma$ molecules are given in Sections \ref{app:1S-states} and \ref{app:2S-states} of the Appendix respectively. They are functions of the relevant angular momentum quantum numbers, and are proportional to the three molecular constants, $\alpha_K$, given by Eq.~(\ref{EqApp:alpha}). The $\alpha_K$ express the size of the scalar, vector and tensor polarizabilities, and their values depend on the frequency of the light. The scalar part shifts all the levels within the subspace equally. This shift is $W_0=-\alpha_0 {\mathcal E}_0^2/4 = -\alpha_0 I/(2 c \epsilon_0)$, where $I$ is the intensity of the light. It is convenient to define $\alpha'_K = \alpha_K/(2 c \epsilon_0)$, so that the scalar Stark shift is simply $W_0=-\alpha'_0 I$.

The vector and tensor polarizabilities produce different shifts for different states, leading to state-dependent trapping potentials. For an angular momentum eigenstate $\ket{j,m_j}$, the expectation value of ${\mathcal A}^1_0$ is proportional to $m_j$. The effect of this vector part can be large when the detuning of the light from a molecular transition is small, or comparable to, the fine-structure splitting of the transition. At larger detunings, the value of $\alpha_1$ is proportional to the ratio of the fine-structure interval to the transition energy, as can be seen from Eq.~(\ref{EqApp:alpha1Approx}). This ratio is normally small, so suppresses the vector part. The value of $\alpha_1$ is also proportional to $\omega_{\rm L}$, so goes to zero when $\omega_{\rm L}=0$. When the light is linearly polarized, all components of ${\mathcal P}^1$ are zero, so the vector polarizability contributes nothing to $H_{\rm S}$. When the light is circularly polarized, the vector part has the same effect as a magnetic field applied along the axis of circular polarization, so can be suppressed by applying a magnetic field orthogonal to that axis \cite{Kaufman2012, Thompson2013}.  As shown by Eq.~(\ref{EqApp:alpha2}), the value of $\alpha_{2}$ is proportional to the difference between the polarizabilities parallel and perpendicular to the molecular bond. These are typically very different, leading to a large value for $\alpha_2$. This tensor part results in trap potentials that depend strongly on the state of the molecule and on the polarization of the light. In Sections~\ref{sec:simple-molecule} and \ref{sec:real-molecule}, we show how to minimize the problems associated with these state-dependent potentials. 

\section{Simple molecule}
\label{sec:simple-molecule}

We first consider a simple diatomic molecule that has no electronic or nuclear spin. We concentrate on the first rotationally-excited state, $N=1$, within the ground electronic state. Excitation from this state to an electronically-excited state with $N=0$ is rotationally closed, as needed for the optical pumping step of the sideband cooling. In this case, we are interested only in the three $m_N$ states of $N=1$, where $m_N$ is the projection of the rotational angular momentum onto a laboratory-fixed $Z$ axis. To make the link with the $^2\Sigma$ molecule considered later, we allow the states $m_N=-1,0,1$ to be non-degenerate at zero intensity, with energies $-w$, 0 and $w$ respectively. 

\subsection{Linearly polarized light}

Our coordinate system is shown in Fig.~\ref{fig:coords}. The trapping light propagates along the $x$ axis. We first assume the light is linearly polarized along the $z$ axis, making an angle $\beta$ to $Z$. The effective Hamiltonian for this system is
\begin{align}
&H_{\rm simple} =
\left(
\begin{array}{ccc}
 -w  & 0 & 0 \\
 0 & 0 & 0 \\
 0 & 0 & w  \\
\end{array}
\right) - 
\left(
\begin{array}{ccc}
 1 & 0 & 0 \\
 0 & 1 & 0 \\
 0 & 0 & 1 \\
\end{array}
\right) \alpha'_0 I - \nonumber \\
&    \left(
\begin{array}{ccc}
 -\frac{1+ 3\cos 2\beta}{20}  & \frac{3 \sin 2\beta }{10 \sqrt{2}} & -\frac{3\sin^2 \beta}{10}  \\
 \frac{3 \sin 2\beta }{10 \sqrt{2}} & \frac{1+ 3 \cos 2\beta}{10} & -\frac{3 \sin 2\beta }{10 \sqrt{2}} \\
 -\frac{3\sin^2 \beta}{10}  & -\frac{3 \sin 2\beta }{10 \sqrt{2}} & -\frac{1+ 3\cos 2\beta}{20}  \\
\end{array}
\right) \alpha'_2 I.
\end{align}
The first matrix gives the energies in the absence of the light, the second is the scalar Stark shift, and the third is the tensor part of the Stark interaction. There is no vector part because our model system has no spin, and because the light is linearly polarized. In terms of the eigenvalues of $H_{\rm simple}$, which we may write as $E(\alpha'_0 I,\alpha'_2 I)$, the scalar Stark shift is $E(\alpha'_0 I,0)-E(0,0)$ and the tensor Stark shift is $E(0,\alpha'_2 I)-E(0,0)$.

\begin{figure}[tb]
    \includegraphics[width=\columnwidth]{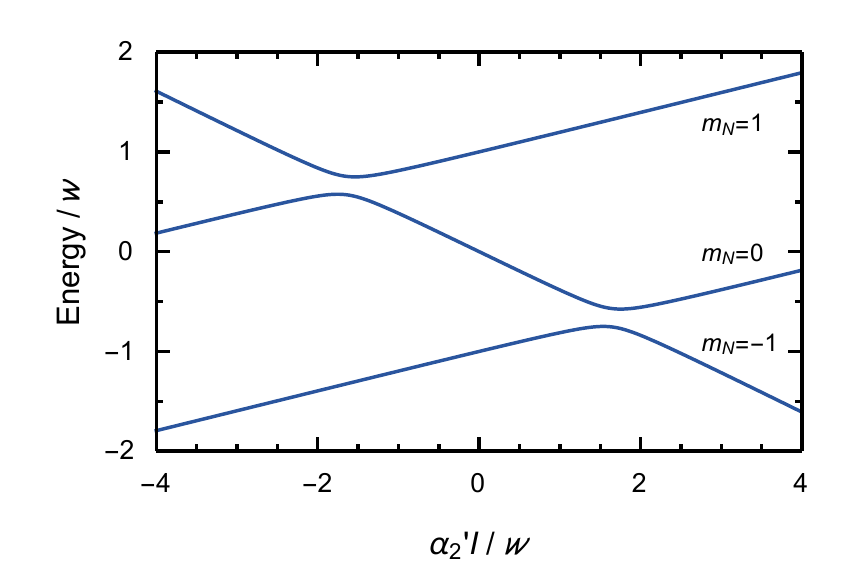}
    \caption{Energies (in units of $w$) of the $N=1$ states of the model molecule as a function of $\alpha'_2 I / w$, in the case where $\beta = \pi/24$. The horizontal axis is proportional to the intensity of the light and to the tensor polarizability of the molecule. The scalar Stark shift, which shifts all states equally, has been subtracted.}
    \label{fig:three_level_energies}
\end{figure}

We can learn a great deal from this simple Hamiltonian. When $w = 0$ the Stark shifts are independent of $\beta$, and when $\beta = 0$ the Stark shifts are independent of $w$. In both cases, the $m_{N}=\pm 1$ states have equal Stark shifts of $\delta E_{\pm 1} = -(\alpha'_0 - \alpha'_2/5)I $, while the $m_N = 0$ state shifts by $\delta E_0 = -(\alpha'_0 + 2\alpha'_2/5)I$. When $\beta = 0$ and $\alpha'_2/w$ is positive (negative), the $m_N=-1 (+1)$ and $m_N=0$ states cross at the intensity where $3/5 \alpha'_2 I = w$ ($-w$). This becomes an avoided crossing when $\beta \ne 0$, and the size of the gap at the avoided crossing is $w\sin(2\beta) /\sqrt{2}$. These features can be seen in Fig.~\ref{fig:three_level_energies}, where we plot the energies of the three states as a function of $\alpha'_2 I/w$ for the case where $\beta = \pi/24$, chosen here and later to clearly highlight the presence of the avoided crossings. We have removed the scalar Stark shift, since it shifts all states equally. We note that the Stark shifts cease to be linear in intensity near the avoided crossing, and that this non-linearity may translate into anharmonicity of the trapping potential if the trap intensity is in this range.

\begin{figure}[tb]
    \includegraphics[width=\columnwidth]{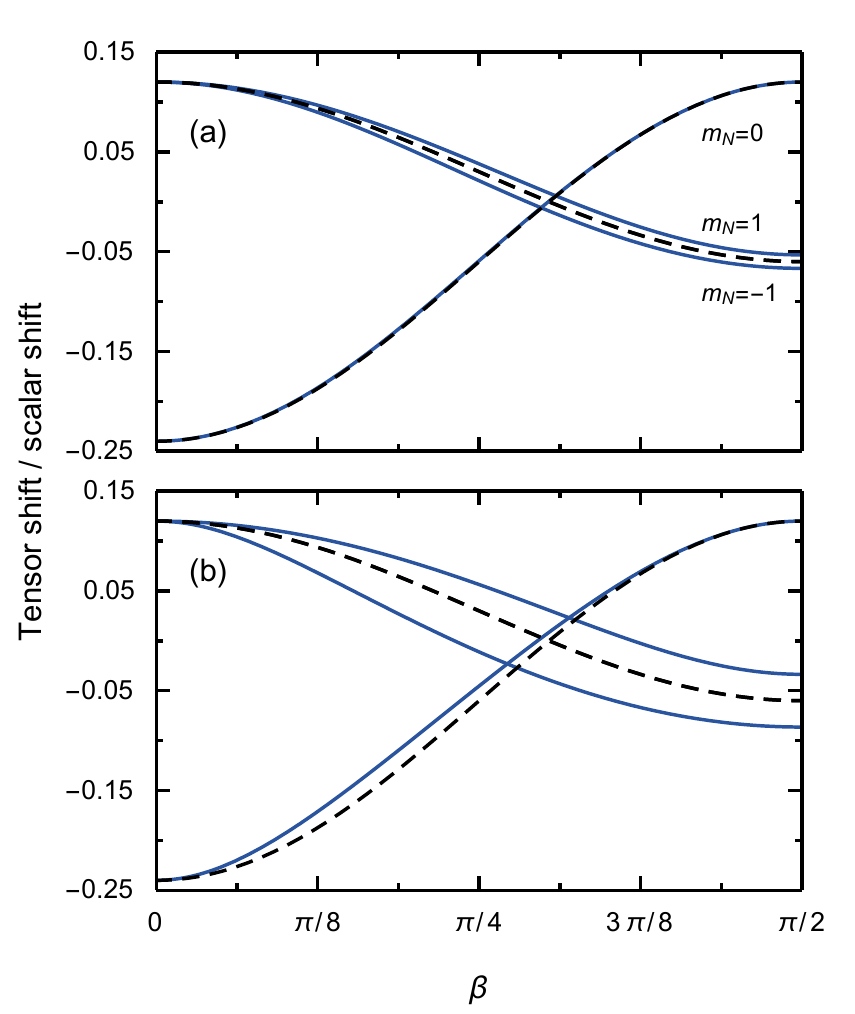}
    \caption{Ratio of tensor Stark shift to scalar Stark shift for the $N=1$ states of the model molecule, as a function of polarization angle $\beta$ from the $Z$ axis. In all cases, $\alpha'_2=-0.6\alpha'_0$. Dashed lines show the case where $|w/(\alpha'_2 I)| \gg 1$. Solid, colored lines are for (a) $w = 4\alpha'_2 I$, (b) $w = \alpha'_2 I$.}
    \label{fig:three_level_shifts}
\end{figure}

Figure \ref{fig:three_level_shifts} shows the ratio of the tensor Stark shift to the scalar Stark shift, as a function of the polarization angle $\beta$. We have chosen $\alpha'_2=-0.6\alpha'_0$, and explore various values of $w$. The dashed lines show the limiting case where $|w/(\alpha'_2 I)| \gg 1$. In this case, the eigenvalues of $H_{\rm simple}$ are very nearly equal to its diagonal elements. By inspection of these elements, we see that the $m_{N}=\pm 1$ states have identical tensor Stark shifts for all values of $\beta$, whereas the tensor Stark shift of $m_{N}=0$ is twice as large and has the opposite sign. We also see that the tensor Stark shift is zero for all three states at the ``magic angle'' where $\beta = \beta_{\rm magic} = \cos^{-1}(1/\sqrt{3})$. The solid lines in Fig.~\ref{fig:three_level_shifts}(a) show the case where $w/(\alpha'_2 I) = 4$. The results follow the dashed lines closely, but the $m_N=\pm 1$ no longer have identical Stark shifts when $\beta \ne 0$. This difference increases as $w$ decreases, as can be seen in Fig.~\ref{fig:three_level_shifts}(b) which shows the case of $w/(\alpha'_2 I) = 1$. In particular, we note that there is no longer any angle where all three states have the same tensor shift.

\subsection{Elliptically polarized light}

\begin{figure}[tb]
    \includegraphics[width=\columnwidth]{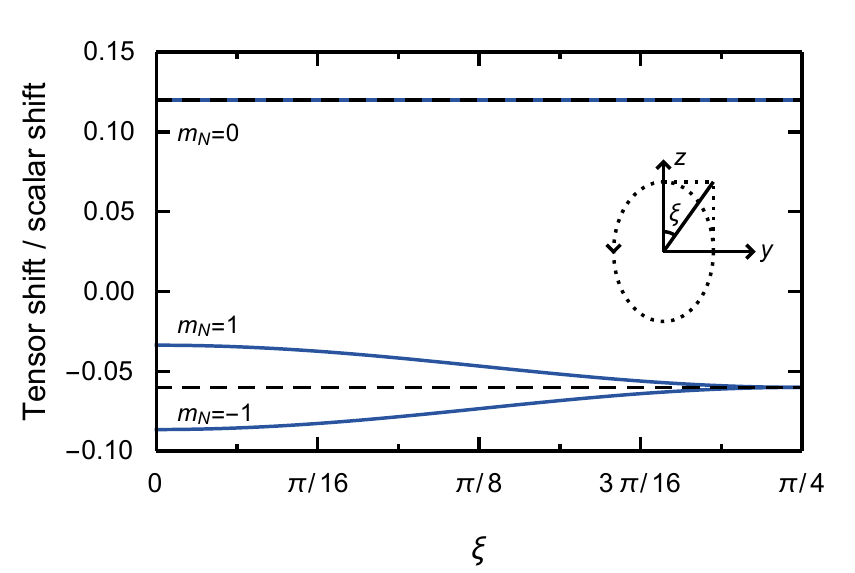}
    \caption{Ratio of tensor Stark shift to scalar Stark shift for the $N=1$ states of the model molecule, as a function of ellipticity parameter $\xi$, defined in the text and shown schematically in inset. $\xi = 0$ corresponds to linear polarization, and $\xi=\pi/4$ to circular polarization. The light propagates along $Z$ and we have chosen $\alpha'_2=-0.6\alpha'_0$. Dashed lines show the case where $|w/(\alpha'_2 I)| \gg 1$. Solid, colored lines are for $w = \alpha'_2 I$.}
    \label{fig:three_level_shifts_elliptical}
\end{figure}

Next we consider the case of elliptically polarized light. The light again propagates along $x$ and $Z$ is aligned with this axis also. The polarization of the light is described by $\hat{\epsilon} = \cos (\xi) \hat{z} -  i \sin (\xi) \hat{y}$. The tensor Stark part of $H_{\rm simple}$ now reads
\begin{equation}
\left(
\begin{array}{ccc}
 -\frac{1}{10}  & 0 & \frac{3}{10} \cos(2\xi) \\
 0 & \frac{1}{5} & 0 \\
 \frac{3}{10}  \cos(2\xi) & 0 &  -\frac{1}{10} \\
\end{array}
\right)\alpha'_2 I.
\end{equation}
When the light is circularly polarized ($\xi = \pm \pi/4$), this matrix is diagonal and the tensor Stark shifts are half the size and of opposite sign relative to the case of light linearly polarized along $Z$. Figure \ref{fig:three_level_shifts_elliptical} shows the ratio of tensor to scalar Stark shifts as a function of $\xi$, with $\alpha'_2=-0.6\alpha'_0$. Once again, the dashed lines show the case where $|w/(\alpha'_2 I)| \gg 1$. In this case, the shifts are independent of $\xi$ and the $m_N=\pm 1$ states shift equally. The solid lines show the case where $w/(\alpha'_2 I) = 1$. The shift of the $m_N=0$ state has no dependence on $\xi$, while the shifts of the $m_N=\pm 1$ states depend on $\xi$ and are different to one another. This difference is largest at $\xi = 0$ (linearly polarized along $z$), zero at $\xi=\pi/4$ (circularly polarized), and reduces as $|w/(\alpha'_2 I)|$ increases.

\subsection{Sideband cooling of simple molecule}

For this simple molecule, sideband cooling could be done with any choice of polarization where two of the three states have equal tensor shifts. This ensures that the Raman frequency for transitions between these two states is independent of the motional state, as required. Here, we consider the specific case where the Raman step is between the $m_N=-1$ and $m_N=+1$ states. The  cooling proceeds as follows: (i) optically pump into $m_N=-1$, (ii) drive the Raman transition from $m_N=-1$ to $m_N=+1$ on a red motional sideband, (iii) repeat. The optical pumping should be on the rotationally-closed transition, which excites the molecule to an $N=0$ state. A pair of laser beams, one linearly polarized along $Z$ and the other circularly polarized about $Z$ achieves the desired optical pumping. 

If it is possible to work in a regime where $|w/(\alpha'_2 I)| \gg 1$, the polarization of the tweezer is not important for the Raman step since the trapping potential is always the same for the $m_N=\pm 1$ states. The tweezer polarization is relevant for the optical pumping step due to spontaneous emission to $m_N=0$ for which the trapping potential is, in general, different. The extra heating this produces can be eliminated by choosing the polarization at the magic angle where the trapping potential is identical for all three states. If it is not feasible to work in the regime where $|w/(\alpha'_2 I)| \gg 1$, the tweezer should be linearly polarized along $Z$, or circularly polarized relative to $Z$, so that the trap potential is identical for the $m_N=\pm 1$ states. We emphasize that other configurations are possible using different pairs of states for the Raman step.

\subsection{Heating during optical pumping}

We now evaluate the extra heating that occurs during the optical pumping step when the third state---not used for the Raman transition---has a different ac Stark shift to the other two. We again focus on the case where the Raman step is between the $m_N=-1$ and $m_N=+1$ states and the third state is $m_N=0$. The $N=0$ excited state has equal branching ratios to each of the three ground states and so it takes, on average, three scattered photons for the molecule to reach the dark state; two of these leave the molecule in $|m_N|=1$ and one in $m_N=0$. The heating that each of these scattered photons produces is given by the three terms in Eq.~(\ref{Eq:scatter-heating}). Let us consider each in turn. The recoil heating, Eq.~(\ref{eq:theta-recoil}), depends on the direction of the absorbed photon and the angular distribution of the emitted photon. In our case, with a single excited state, specifying the initial and final value of $|m_N|$ is sufficient to uniquely define both the beam from which the photon is absorbed and the angular distribution of the emitted photon. By analogy with Eq.~(\ref{eq:av-recoil-energy}) we define $\myol{E}_{\rm rec}^{i,j}$, the average recoil energy for a scattering event which takes the molecule from a ground state with $|m_N|=i$ to a ground state with $|m_N|=j$. The circularly polarized beam which couples to $m_N=+1$---from which on average two photons are scattered---is necessarily parallel to the $\mathcal{B}$ field, but the linearly polarized beam which couples to the $m_N=0$ state---from which on average one photon is scattered---can propagate along any direction perpendicular to that. We can control the heating along a particular trapping axis to some extent by choosing this angle appropriately. It is likely to be helpful to choose it orthogonal to the optical axis of the trapping light where, as we will see in Sec.~\ref{sec:real-tweezer}, the confinement is weakest. 

The contribution from $\Theta_{\rm disp}$, Eq.~(\ref{eq:theta-disp}), is zero because the potentials are not displaced with respect to one another. To understand the contribution of $\Theta_{\rm curv}$, given by Eq.~(\ref{eq:theta-curv}), we need to know how many of the scattering events change $|m_N|$. Let $y_{m_N}$ be the mean number of $|m_N|$-changing events needed to reach the dark state, starting from state $m_N$. Consider a molecule initially in $m_N=-1$. If it scatters a photon and decays to $m_N=1$, the dark state is reached with no $|m_N|$-changing events. If it decays to $m_N=0$, there has been one $|m_N|$-changing event, and there are an average of $y_0$ more to come. If it decays back to $m_N=-1$, there are an average of $y_{-1}$ events to come. Each outcome has a probability of 1/3. Thus
\begin{equation}
    y_{-1} = \frac{1}{3}y_{-1} + \frac{1}{3}(1+y_{0}).
\end{equation}
By a similar argument
\begin{equation}
    y_{0} = \frac{1}{3}(1+y_{-1}) + \frac{1}{3}y_{0} + \frac{1}{3}.
\end{equation}
Together these equations give $y_{-1}=4/3$ and $y_{0}=5/3$. Since the molecule begins and ends the optical pumping step in a state with $|m_N|=1$, half of the $4/3$ events change $|m_N|$ from $1$ to $0$, and half the reverse.

\begin{figure}
    \includegraphics[width=\columnwidth]{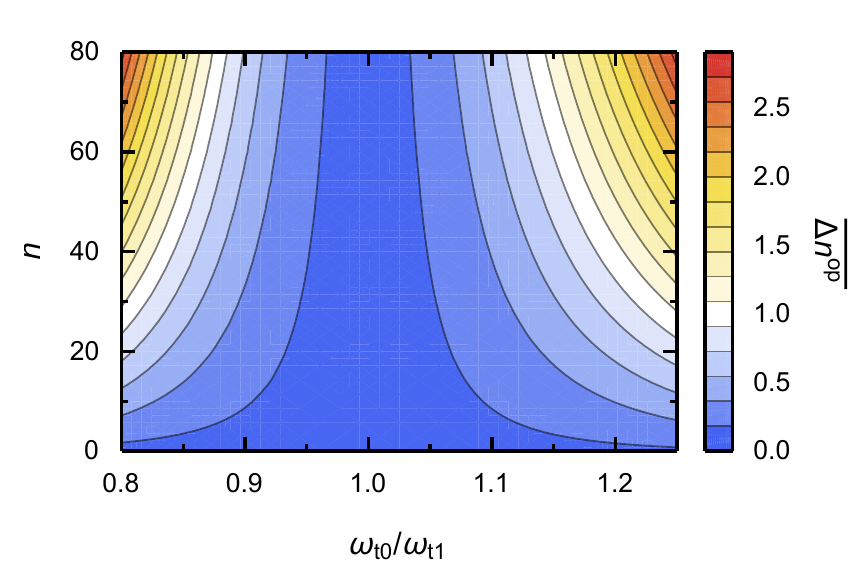}
    \caption{Mean change in $n$ during optical pumping $\overline{\Delta n^{\rm op}}$ as a function of $n$ and $\omega_{\rm t0}/\omega_{\rm t1}$ for $\eta_{1}=0.2$. The circularly polarized optical pumping beam is assumed parallel to the trap axis and the linearly polarized beam perpendicular.\label{fig:heating}}
\end{figure}

We can now use Eq.~(\ref{Eq:scatter-heating}) to estimate the mean change in motional quantum number during optical pumping, $\overline{\Delta n^{\rm op}}$. If $\overline{\Delta n^{\rm sc}_{i,j}}\ll 1$ for all $i,j\in\{-1,0,1\}$, so that we can assume a fixed $n$ in Eq.~(\ref{eq:theta-curv}), then to a good approximation
\begin{equation}
\begin{split}
    \overline{\Delta n^{\rm op}} =& \frac{4}{3}\frac{\myol{E}_{\rm rec}^{1,1}}{\hbar \omega_{\rm t1}}+\frac{2}{3}\frac{\myol{E}_{\rm rec}^{1,0}}{\hbar \omega_{\rm t0}}+\frac{2}{3}\frac{\myol{E}_{\rm rec}^{0,1}}{\hbar \omega_{\rm t1}}+\frac{1}{3}\frac{\myol{E}_{\rm rec}^{0,0}}{\hbar \omega_{\rm t0}}\\
    &\quad+\frac{2}{3}(\frac{\omega_{\rm t1}}{\omega_{\rm t0}}+\frac{\omega_{\rm t0}}{\omega_{\rm t1}} -2)(n+\frac{1}{2}).
    \label{Eq:heating-op}
\end{split}
\end{equation}
Here, $\omega_{\mathrm{t}i}$ is the trap frequency corresponding to the ground states with $|m_N|=i$. Figure~\ref{fig:heating} shows how $\overline{\Delta n^{\rm op}}$ varies as a function of both $n$ and the ratio of the trap frequencies, for $\eta_{1}=0.2$, which is a realistic Lamb-Dicke parameter for trap frequencies near 200~kHz. The blue areas of the plot show the parameter space where $\overline{\Delta n^{\rm op}}<1$. This corresponds to net cooling along the axis being cooled when the coherent step drives the first-order red sideband. We will see later that $\omega_{\rm t1}/\omega_{\rm t0}$ is typically in the range 0.8--1.25. Throughout this range, $\overline{\Delta n^{\rm op}}<1$ all the way up to $n \approx 25$.

\section{Real molecule}
\label{sec:real-molecule}

Next, we consider a real molecule with a $^{2}\Sigma$ ground state and a nuclear spin. We will use CaF as an illustrative example, though our discussion will apply to other molecules of this type. For laser cooling, the electronic transition to either the A$^{2}\Pi_{1/2}$ state or the B$^{2}\Sigma$ state can be used. For sideband cooling, we choose to use the transition ${\textrm B}^{2}\Sigma (v=0, N=0) \leftarrow  {\textrm X}^{2}\Sigma (v=0, N=1)$. Here, $v$ and $N$ refer to the vibrational and rotational quantum numbers respectively.  With this choice of excited state, decay to any other rotational state of X is forbidden by the parity and angular momentum selection rules, so the transition is rotationally closed. The branching ratio for decays to other vibrational states depends on the choice of molecule. For CaF, it is particularly small, about $10^{-3}$, so that for the purpose of sideband cooling we can consider the transition to be vibrationally closed. For other laser-coolable molecules with less favorable branching ratios, vibrational repump lasers can be used.

To understand how to apply sideband cooling, we need to consider the hyperfine interactions in the ground state. For CaF, and similar molecules, the Hamiltonian describing these interactions is
\begin{equation}
H_{\rm hfs} =\gamma \vec{S}\cdot\vec{N} + \left(b+\frac{c}{3}\right) \vec{I}\cdot\vec{S} + \frac{c}{3} \sqrt{6}T^2(C)\cdot T^2(\vec{I},\vec{S}),
\label{eq:Hhfs}
\end{equation}
where $\vec{N}$, $\vec{S}$ and $\vec{I}$ are the dimensionless operators for the rotational angular momentum, electron spin and nuclear spin. The first term is the spin-rotation interaction, while the second and third represent the interaction between the electron and nuclear magnetic moments. Here, $T^2(\vec{I},\vec{S})$ is the rank-2 spherical tensor formed from $\vec{I}$ and $\vec{S}$, while $T^2(C)$ is a spherical tensor whose components are the spherical harmonics $C^2_q(\theta,\phi)$. We have neglected the nuclear-spin-rotation interaction which is much smaller than the other terms. 

\subsection{Reduction to the simplified molecule}

The hyperfine interactions couple together the angular momenta, and as a result the ac Stark shift is, in general, much more complicated than the simple picture described above. However, that simple picture can be recovered by applying a magnetic field, $\vec{\mathcal{B}}$, that is large enough to uncouple the angular momenta. The Zeeman Hamiltonian for a $^{2}\Sigma$ state is
\begin{align}
H_{\rm Z} &= g_{S}\mu_{\rm{B}} \vec{S}\cdot \vec{\mathcal{B}} - g_{\rm{N}}\mu_{\rm{N}}\vec{I}\cdot \vec{\mathcal{B}} - g_{\rm{r}}\mu_{\rm{B}}\vec{N}\cdot \vec{\mathcal{B}} \nonumber \\ &+ g_l \mu_{\rm{B}}(\vec{S} \cdot \vec{\mathcal{B}} - (\vec{S}\cdot \hat{\lambda})(\vec{\mathcal{B}}\cdot \hat{\lambda})),
\label{eq:HZeeman}
\end{align}
where $\hat{\lambda}$ is a unit vector along the internuclear axis, and we have assumed that only one nucleus has a spin. The first term is due to the unpaired electron spin and is typically $10^{3}$ times larger than the other terms. We will often only need to consider this term. When $\vec{\mathcal{B}}$ is large, so that the Zeeman interaction is much larger than the hyperfine interaction, the eigenstates are well described by uncoupled angular momentum eigenstates, $\ket{N,m_N}\ket{S,m_S}\ket{I,m_I}$. Each rotational state splits into two manifolds with $m_{S}=\pm 1/2$, whose Zeeman shifts are $\Delta E_{\rm Z} \approx g_{S}\mu_{\rm B} m_{S} \mathcal{B} \approx \pm \mu_{\rm B} \mathcal{B}$. Here, we have used $g_{S}\approx 2$ and have neglected the small terms. The hyperfine interaction lifts the degeneracy with respect to $m_N$ and $m_I$ within each of these manifolds. In the limit where the angular momenta are completely uncoupled, the ac Stark shift has no dependence on $m_S$ and $m_I$. Furthermore, the values of $m_S$ and $m_I$ cannot change in either the Raman step or the optical pumping step.\footnote{Here, our choice of the B$^{2}\Sigma$ excited state is important: in strong magnetic field, $m_S$ and $m_I$ are good quantum numbers for both the ground and excited states. This would not be the case for the A$^{2}\Pi$ excited state because the fine-structure splitting is large compared to any reasonable Zeeman splitting.} Having chosen a particular ($m_S$, $m_I$) pair, their values are fixed, so that (for $N=1$) we are left with only three states, just as in Sec.~\ref{sec:simple-molecule}. 

Taking $N=1$ and $S=I=1/2$, the shifts due to $H_{\rm hfs}$ to first-order in perturbation theory are
\begin{equation}
\Delta E_{\rm hfs} \approx \gamma m_N m_S + \left(b+\frac{c}{3}\right)m_S m_I + (-1)^{m_N} \frac{4c}{15} \frac{m_S m_I}{m_N^2+1}.
\end{equation}
Relative to $m_N=0$, the energies of the $m_N=\pm 1$ states are
\begin{equation}
w_{\pm} = \pm \gamma m_S - \frac{2c}{5} m_I m_S.
\label{eq:Deltapm}
\end{equation}
In the limit where $c \ll \gamma$, the splitting is symmetric and the description is identical to that of Sec.~\ref{sec:simple-molecule}. For CaF, $\gamma$ and $c$ are almost equal, so the splitting is not quite symmetric, though this makes little difference to the description.

In practice, the magnetic field is limited in strength. This has two important consequences. First, in the optical pumping step, the residual state mixing by $H_{\rm hfs}$ can result in decay to a different manifold of states from the one selected. Second, the ac Stark shifts deviate from the simple behavior shown in Figs.~\ref{fig:three_level_energies}, \ref{fig:three_level_shifts} and \ref{fig:three_level_shifts_elliptical}. Next, we work out the severity of these imperfections to our scheme.

\subsection{Residual state mixing}

Let us write the uncoupled states using the notation $\ket{m_N,m_S,m_I}$. All three terms in $H_{\rm hfs}$ result in mixing of these uncoupled states, and we can calculate the mixing amplitudes by perturbation theory (assuming that $\mu_{\rm B}\mathcal{B} \gg \gamma, b, c$). The spin-rotation interaction, $\gamma \vec{N}\cdot \vec{S}$, has no effect on the upper state of the transition which has $N=0$, but it does change the lower states since they have $N=1$. The state $|{m_N,\pm 1/2,m_I}\rangle$ obtains an admixture of $|{m_N\pm 1,\mp 1/2,m_I}\rangle$ (where that state exists), with amplitude $\pm \gamma/(2\sqrt{2}\mu_{{\rm B}}\mathcal{B})$. It follows that the excited state with $m_{S}=\pm1/2$ can decay to the ground state with (nominally) $m_{S}=\mp1/2$ with a branching ratio of
\begin{equation}
b_{{\rm r},1}=\frac{2}{3}\left(\frac{\gamma}{2\sqrt{2}\mu_{{\rm B}}\mathcal{B}}\right)^2.
\end{equation}
Importantly, this branching ratio is suppressed with increasing $\mathcal{B}$. For CaF at $\mathcal{B}=300$~G, we find $b_{{\rm r},1}=7.4\times 10^{-4}$. Similarly, due to the $\vec{I}\cdot\vec{S}$ term of Eq.~(\ref{eq:Hhfs}), the state  $|{m_N,\pm 1/2,\mp 1/2}\rangle$ obtains an admixture of $|{m_N,\mp 1/2,\pm 1/2}\rangle$ with an amplitude $(b+c/3)/(4 \mu_{{\rm B}}\mathcal{B})$. This affects both the X and B states. The B state with $m_{S}=-m_{I}=\pm1/2$ can decay to an X state with the opposite values of $m_{S}$ and $m_{I}$ with a branching ratio of
\begin{equation}
b_{{\rm r},2}=(\epsilon_{{\rm X}} +\epsilon_{{\rm B}})^2,
\end{equation}
where
\begin{equation}
\epsilon_{{\rm X/B}} = \left(\frac{b+c/3}{4 \mu_{{\rm B}}\mathcal{B}}\right)_{{\rm X/B}},
\end{equation}
and the subscript indicates which state the hyperfine coefficients belong to. For CaF at $\mathcal{B}=300$~G, we find $b_{{\rm r},2}=7.2\times 10^{-3}$. This decay route can be eliminated by choosing to use a manifold with $m_S=m_I$.

\begin{figure}[tb]
    \includegraphics[width=\columnwidth]{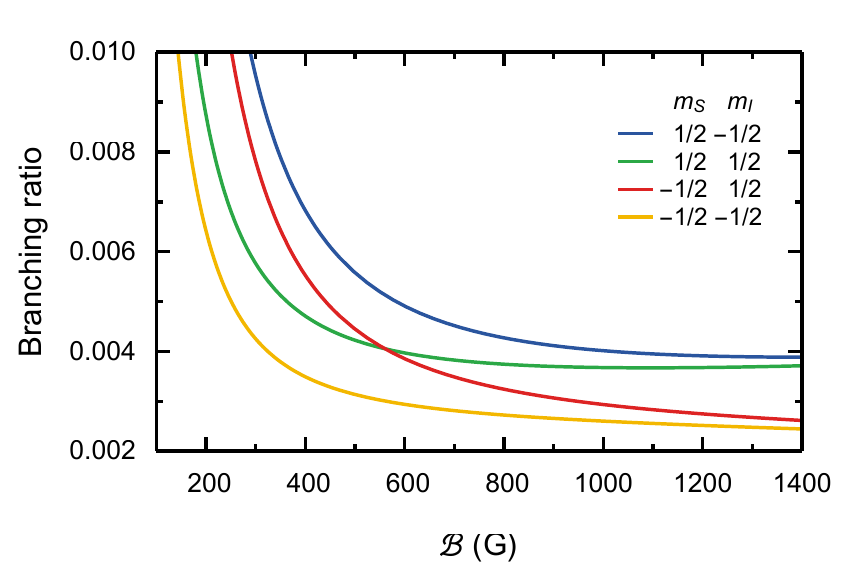}
    \caption{Branching ratio as a function of magnetic field for decay from each spin manifold of the B$^2\Sigma^+ (N=0)$ states of CaF to a different spin manifold within X$^2\Sigma^+(N=1)$. In order of increasing branching ratio at $\mathcal{B}=\SI{800}{\gauss}$ the states have $(m_S,m_I)=(-1/2,-1/2)$, $(-1/2,1/2)$, $(1/2,1/2)$, $(1/2,-1/2)$.  We have set the light intensity to zero.
    \label{fig:branching}}
\end{figure}

Next, consider the last term of Eq.~(\ref{eq:Hhfs}). In the uncoupled basis, it has non-zero matrix elements between all pairs of states with equal $m_{F}=m_N+m_S+m_I$. This means that as well as coupling states that differ in $m_S$, it couples states of the same $m_S$ that differ in $m_N$ and $m_I$. The former couplings are suppressed by the large Zeeman splitting between opposite $m_S$ manifolds, in the same way as for the first two terms of Eq.~(\ref{eq:Hhfs}) discussed above, but the latter couplings are not suppressed by the field because the terms in the Zeeman Hamiltonian that depend on $m_N$ and $m_{I}$ are very small. As an example, consider the nominal state $\ket{-1,-1/2,1/2}$. In perturbation theory, its admixture with $\ket{-1,1/2,-1/2}$ has amplitude $-c/(60 \mu_{\rm B}\mathcal{B})$. This can cause $m_{S}$ to change in the excited state decay. For CaF, the branching ratio is smaller than the terms already discussed above, because $c$ and $\gamma$ are approximately equal. Our chosen nominal state also has an admixture with $|0,-1/2,-1/2\rangle$. A rough estimate of its amplitude can be obtained by treating the last term in $H_{\rm hfs}$ as a perturbation to the other two terms, giving an amplitude
\begin{equation}
\zeta \approx -\frac{c}{\sqrt{2}(5\gamma - 5b - 5c/3)}.\nonumber
\end{equation}
Similarly, using the same approximation, the nominal state $|0,-1/2,1/2\rangle$ has an admixture with $|1,-1/2,-1/2\rangle$ with amplitude $-\zeta$. It follows that the B$^2\Sigma$ state with $m_{S}=m_{I}=-1/2$ can decay to the nominal X states $|m_N,-1/2,1/2\rangle$ with $m_N=-1,0$, with a branching ratio of $b_{{\rm r},3} \approx 2/3 \zeta^2.$
The situation is the same for the other three states of B$^2\Sigma$---in each case there is a leak out of the chosen $(m_S,m_I)$ manifold of approximate size $b_{{\rm r},3}$ which cannot be reduced by applying a magnetic field of any reasonable size. For CaF, $b_{{\rm r},3} \approx 3 \times 10^{-3}$.  

Finally, we note that there is another mechanism for mixing states of different $m_N$ and $m_I$ but the same $m_S$. This is through the combination of the $\vec{S}\cdot\vec{N}$ and $\vec{I}\cdot\vec{S}$ terms. The amplitude for this is proportional to the product of two matrix elements, one for each term, but scales only as $1/\mathcal{B}$ because the mixing is with a state from the same $m_S$ manifold through an intermediate state of opposite $m_S$. This mechanism affects 8 out of the 12 states of X and can be just as strong as the more direct mechanisms discussed above.

Figure \ref{fig:branching} shows the exact branching ratio, calculated numerically, for each of the spin manifolds of the B$^2\Sigma^+ (N=0)$ state of CaF to decay to a different spin manifold of X$^2\Sigma^+(N=1)$, as a function of $\mathcal{B}$. The behavior is as discussed above---the branching ratios scale as $1/\mathcal{B}$ towards a constant value that is close to $b_{{\rm r},3}$. The branching ratios depend on the choice of spin manifold, and we see that using the $(m_S,m_I)=(-1/2,-1/2)$ manifold minimizes the leak to other manifolds.

\subsection{Tensor Stark shifts}

\begin{figure}[tb]
    \includegraphics[width=\columnwidth]{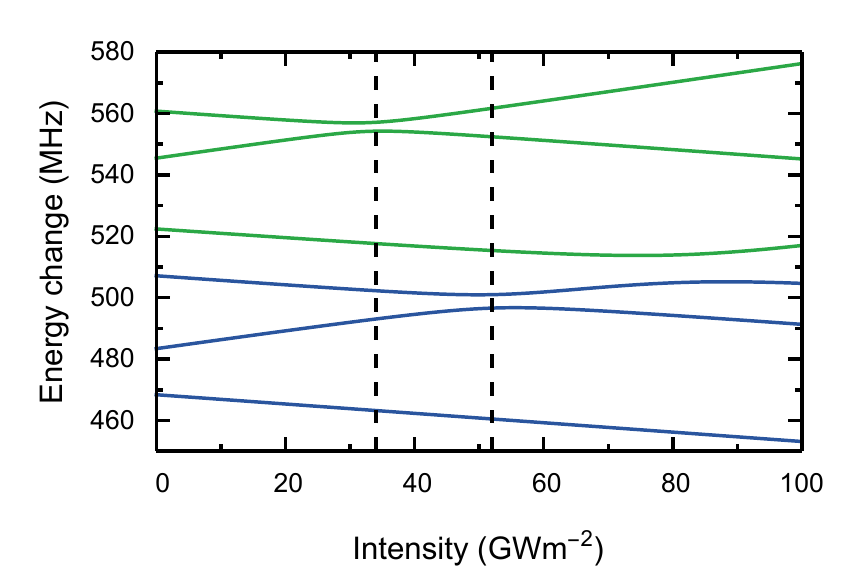}
    \caption{Energy levels of CaF as a function of light intensity, in a magnetic field of 300~G. All levels have $N=1, m_S=1/2$. For clarity, the rotational energy and the scalar Stark shift, which are common to all the levels, have been subtracted. The light is linearly polarized at an angle of $\pi/24$ to the magnetic field axis. In ascending order of energy, the states have $(m_N,m_I)=(-1,-1/2)$, $(0,-1/2)$, $(1,-1/2)$, $(-1,1/2)$, $(0,1/2)$, $(1,1/2)$. The color of the lines indicates $(m_S,m_I)$ and follows the same scheme used in Fig.~\ref{fig:branching}. Dashed lines show the intensities where the three-level model predicts avoided crossings between $m_N = 1$ and $m_N=0$ levels. The pattern of levels with $m_S=-1/2$ is almost identical, but with states in the opposite order.}
    \label{fig:CaF_levels}
\end{figure}

Next, we calculate the tensor Stark shifts of CaF molecules in the presence of a strong magnetic field, and compare the results to those of the three-level model presented in Sec.~\ref{sec:simple-molecule}. We calculate the eigenvalues of $H_{\rm tot} = H_{\rm Stark}+H_{\rm hfs}+H_{\rm Z}$ given by Eqs.~(\ref{Eq:HStark}), (\ref{eq:Hhfs}) and (\ref{eq:HZeeman}). We suppose the optical trap has a wavelength of 780~nm, and estimate the values of $\alpha_K$ by assuming that the A$^2\Pi$ and B$^2\Sigma^+$ states dominate the sums over states in Eq.~(\ref{EqApp:alphaSumStates}). The energies of the states are calculated using the molecular constants given in \cite{Kaledin1999}, and the dipole moments using the data given in \cite{Childs1984, Wall2008, Dagdigian1974}. We find $\alpha_0' \approx 1.4 \times 10^{-3}$~Hz/(W/m$^2$), $\alpha_1' \approx 3 \times 10^{-5}$~Hz/(W/m$^2$) and $\alpha_2' \approx -8 \times 10^{-4}$~Hz/(W/m$^2$). 

\begin{figure*}[t]
    \includegraphics[width=\textwidth]{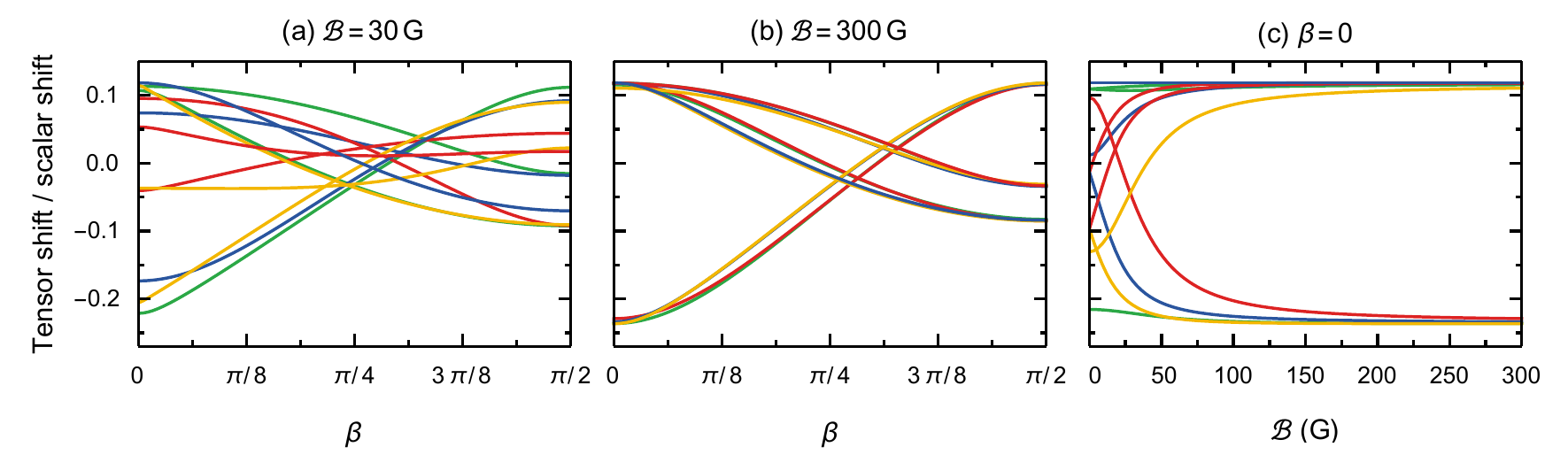}
    \caption{Ratio of tensor Stark shift to scalar Stark shift for the $N=1$ states of CaF. (a,b) As a function of polarization angle $\beta$ with magnetic field of (a) 30~G (b) 300~G. (c) As a function of magnetic field with polarization parallel to magnetic field. The light intensity is $I=25$~GW/m$^2$ throughout. The color of the lines indicates the value of $(m_S,m_I)$ and follows the same scheme used in Fig.~\ref{fig:branching}.}
    \label{fig:CaF_shifts}
\end{figure*}

Figure \ref{fig:CaF_levels} shows the eigenvalues of $H_{\rm tot}$, focusing on the levels that have $N=1$ and $m_S=1/2$. We have chosen $\mathcal{B}=300$~G and linearly polarized light at angle $\beta=\pi/24$, the same as used for Fig.~\ref{fig:three_level_energies}. The upper three levels have $m_{I}=1/2$, and their shifts with intensity are similar to those in Fig.~\ref{fig:three_level_energies} (remembering that $\alpha_2'I/w$ is negative). The lower three levels have $m_{I}=-1/2$ and again show similar shifts with intensity. Our three-level model predicts an avoided crossing between $m_N=1$ and $m_N=0$ at an intensity $I_{\rm c} = -5/3 w_{+}/\alpha_2'$, where $w_{+}$ is given by Eq.~(\ref{eq:Deltapm}). These values are $I_{\rm c}=34$ and \SI{52}{\giga\watt\per\meter\squared} for $m_I=1/2$ and $-1/2$ respectively. These intensities are indicated by the dashed lines in Fig.~\ref{fig:CaF_levels}, and we see that the avoided crossings do indeed occur very close to these values. At intensities close to $I_{\rm c}$ the trapping potential will be distorted due to the non-linearity of the Stark shift with intensity around the avoided crossing. We note that, for CaF, with $m_S=m_I=1/2$, $\alpha_0' I_{\rm c} h/k_{\rm B} = \SI{2.1}{\milli\kelvin}$.

Figure \ref{fig:CaF_shifts}(a) shows the ratio of the tensor Stark shift to the scalar Stark shift for the $N=1$ levels of CaF, at an intensity of $I=\SI{25}{\giga\watt\per\meter\squared}$, a magnetic field of $\mathcal{B}=30$~G, and as a function of polarization angle $\beta$. Every level has a different Stark shift and a different dependence on $\beta$. Figure \ref{fig:CaF_shifts}(b) shows the same information for $\mathcal{B}=300$~G, showing that at this higher field the levels group together and the pattern of shifts resembles the simple one shown in Fig.~\ref{fig:three_level_shifts}. Our chosen intensity gives $\alpha_2'I = -19.3$~MHz, which corresponds to $\alpha_2'I/w_+ = -1.22$ for positive $m_S m_I$, and $\alpha_2'I/w_+ = 0.81$ for negative $m_S m_I$, with $w_+$ given by Eq.~(\ref{eq:Deltapm}). Thus, at high $\mathcal{B}$, we expect a close resemblance to Fig.~\ref{fig:three_level_shifts}(b), which is indeed what we see in Fig.~\ref{fig:CaF_shifts}(b). The small splitting of the three curves into closely-spaced pairs is due to the different values of $w_+$ for opposite signs of $m_S m_I$. Figure \ref{fig:CaF_shifts}(c) shows the ratio of the tensor Stark shift to the scalar Stark shift for $\beta=0$ as a function of $\mathcal{B}$. At fields approaching \SI{300}{\gauss} the ac Stark shifts of the 12 states separate into two groups corresponding to states with $|m_N|=1$ and $|m_N|=0$ as expected.

\section{Real light: a tweezer trap}
\label{sec:real-tweezer}

Real tweezer traps are produced using a high numerical aperture (NA) lens to focus light down to a spot size comparable to the wavelength. To model a real trap, we use the vector Debye integral \cite{Richards1959} to compute the distribution of intensity and polarization close to the focus of a trap with parameters suitable for CaF. We then find the trap potential for all the $N=1$ states of the molecule by calculating the eigenvalues of $H_{\rm tot}$ at each point in the distribution. The calculations are for a \SI{780}{\nano\meter} input beam propagating along $x$ and linearly polarized along $z$, focused through a lens of 0.55 NA. The $1/e^2$ diameter of the input beam is equal to the lens diameter and the total power is \SI{20}{\milli\watt}. These parameters give a peak intensity at the center of the tweezer of \SI{25}{\giga\watt\per\meter\squared}. For CaF molecules in the $N=1$ manifold with a \SI{300}{\gauss} $\mathcal{B}$ field applied parallel to the incident polarization vector, the trap frequencies are nearly equal (less than 1\,\% difference) for states with the same value of $|m_N|$. For a molecule in $m_N=\pm1$(0), we calculate trap frequencies of \SI{213(174)}{\kilo\hertz} parallel to the incident polarization, \SI{224(187)}{\kilo\hertz} perpendicular to both the incident polarization and the optical axis, and \SI{38(32)}{\kilo\hertz} parallel to the optical axis.

The intensity profile of the trap is not perfectly harmonic and the tight focusing gives rise to polarization gradients close to the focus which can further distort the trap shape. Here, we consider the effects of these imperfections.

\subsection{Anharmonicity}

The tweezer trap potential is anharmonic away from the trap center due to the approximately Gaussian intensity profile of the trap. Further distortions of the potential are introduced when the intensity is close to an avoided crossing between the internal states of the molecule (see Fig.~\ref{fig:CaF_levels}). Any anharmonicity causes the sideband frequencies to depend on the motional state of the molecule.

At intensities far from any avoided crossings, we find the potential near the trap center is approximately harmonic with a small, negative, quartic perturbation. The motional Hamiltonian is
\begin{equation}
    H_\mathrm{t} = \frac{p^2}{2 M} + \frac{1}{2}M\omega^2 z^2 - f z^4.
\end{equation}
Working in the natural units of the system with $\tilde{z}= z \sqrt{M\omega/\hbar}$, $\tilde{p}= p /\sqrt{\hbar M \omega}$, the dimensionless motional Hamiltonian, $\tilde{H_\mathrm{t}}= H_\mathrm{t}/(\hbar \omega)$, can be written
\begin{equation}
    \tilde{H_\mathrm{t}} = \frac{1}{2}\tilde{p}^2 + \frac{1}{2} \tilde{z}^2 - \tilde{f} \tilde{z}^4,
\end{equation}
where $\tilde{f} = (\hbar/(M^2 \omega^3)) f \ll 1$. First order perturbation theory gives the dimensionless energy, $\tilde{E}_n = E_n/(\hbar \omega)$, of the $n$'th motional eigenstate as
\begin{equation}
\begin{split}
    \tilde{E}_n &\simeq n + \frac{1}{2} -\tilde{f}\bra{n}\tilde{z}^4\ket{n},\\
    &= n + \frac{1}{2} -\frac{\tilde{f}}{4}(3 + 6n + 6n^2).
\end{split}
\end{equation}
The anharmonicity is apparent from the $n$ dependence of the trap level spacing,
\begin{equation}
    \tilde{E}_{n+1} - \tilde{E}_n = 1 - 3\tilde{f} - 3\tilde{f} n.
\end{equation}
The $N=1$ states of CaF in a \SI{300}{\gauss} field have $\tilde{f} \sim\num{7e-4}$ in the radial direction and $\sim\num{1e-4}$ in the axial direction, meaning the level spacing changes by less than 5\,\% for motional states up to $n \sim 25$ and $n \sim 170$ respectively. This effect is negligible, provided the effective Rabi frequency for the Raman process is not chosen too small compared to the trap frequency.

\subsection{Polarization gradients}

\begin{figure}
    \includegraphics[width=\columnwidth]{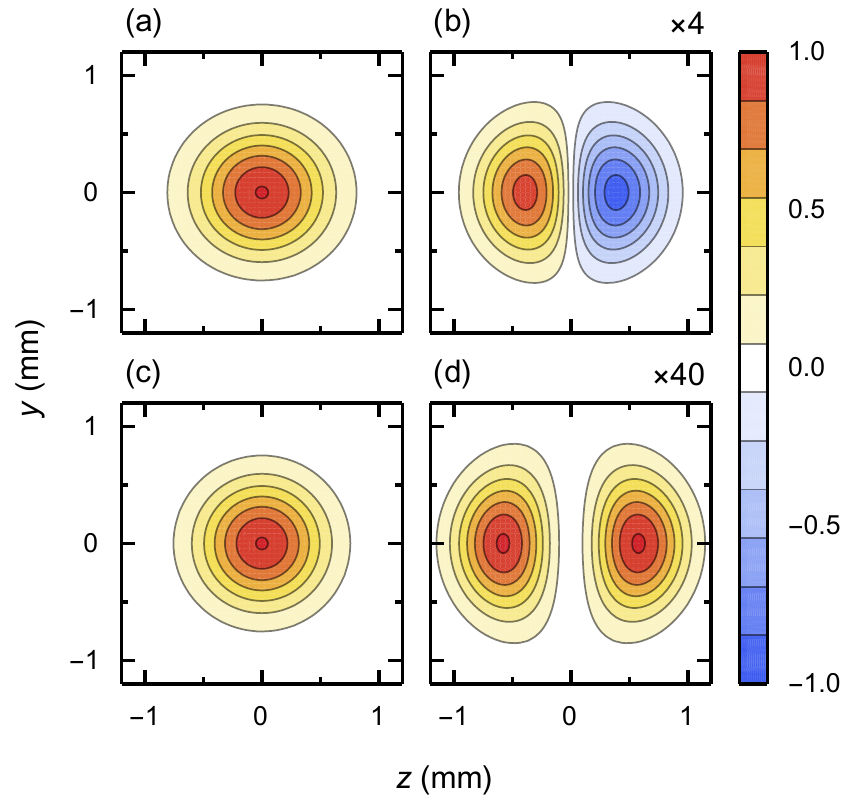}
    \caption{Calculated elements of the $\mathcal{S}$ tensor, defined in the text. (a) $\mathcal{S}_{0}^0$, (b) $\frac{1}{\sqrt{2}}\mathrm{Im}(\mathcal{S}_{-1}^1+\mathcal{S}_{1}^1)$, (c) $\mathcal{S}_{0}^2$, (d) $\frac{1}{\sqrt{2}}\mathrm{Re}(\mathcal{S}_{-2}^2+\mathcal{S}_{2}^2)$. The elements shown in (b) and (d) have been scaled by factors of 4 and 40 respectively in order to be shown on the same scale.}
    \label{fig:light-field}
\end{figure}

Figure~\ref{fig:light-field} shows example elements of the polarization tensor at positions across the focal plane, multiplied by the ratio of the local intensity to the maximum intensity at the trap focus: $\mathcal{S}=(I/I_{\rm max})\mathcal{P}$. The elements are evaluated with respect to the incident polarization vector. Figure~\ref{fig:light-field}(a) shows the scalar component $\mathcal{S}^0_0$. As shown in Eq.~(\ref{SEq:PTensor00}), $\mathcal{P}^0_0$ is 1 everywhere and so the plot mirrors the Gaussian intensity profile of the light with a waist of \SI{0.74}{\micro\meter}. Parts (b)-(d) of Fig.~\ref{fig:light-field} illustrate how the non-paraxial focusing of the light creates polarization gradients across the trapping volume. The polarization tensor is, in general, complex, but useful real quantities can be obtained from linear superpositions of elements, as in Fig.~\ref{fig:light-field}.

For displacements away from the focus along the polarization vector of the incident light, the local polarization has a circular component with handedness along $y$, perpendicular to both the incident polarization vector and the optical axis. The quantity $\frac{1}{\sqrt{2}}\mathrm{Im}(\mathcal{S}_{-1}^1+\mathcal{S}_{1}^1)$, shown in Fig.~\ref{fig:light-field}(b), is proportional to the intensity of circularly polarized light along this axis. The component has opposite handedness on either side of the focus, reflected by the change of sign in the plot, and couples to the vector part of the polarizability. Within a given hyperfine level, the vector contribution to the Hamiltonian looks like a fictitious magnetic field along the axis of the circular component. The gradient of polarization creates a gradient of this fictitious magnetic field that can shift the center of the trap for different $m_F$ states. The effect can be reduced to a negligible level by applying a large magnetic field orthogonal to the fictitious magnetic field, as previously demonstrated for atoms~ \cite{Kaufman2012, Thompson2013}. Our method for sideband cooling of molecules already requires this large applied field, so the suppression will be automatic. 

Figure~\ref{fig:light-field}(c) shows component $\mathcal{S}^2_0$. It is equal to 1 at the trap center where the light field is linearly polarized. The structure is very similar to that of the intensity distribution in Fig.~\ref{fig:light-field}(a) with a slight and asymmetric narrowing caused by the polarization gradient. The component couples to the tensor polarizability to provide a second mechanism by which the vector contribution is suppressed. To see this, choose a quantization axis for the molecule along $z$---the incident polarization direction. The effect of the vector Stark shift is that of a magnetic field orthogonal to this axis, introducing an off-diagonal matrix element---let's call it $a_1$---that couples $m_N=\pm 1$ to $m_N=0$. The $\mathcal{P}^2_0$ component of the polarization tensor, introduces diagonal matrix elements which shift the $m_N=\pm 1$ states relative to $m_N=0$ by an amount $a_2$. When $a_2 \gg a_1$, as is often the case in molecules where the tensor polarizability is large compared to the vector polarizability, the coupling $a_1$ is ineffective because it couples states far apart in energy; the effect of the vector part is suppressed relative to $a_1$ by the factor $a_1/a_2$.

Figure~\ref{fig:light-field}(d) shows $\frac{1}{\sqrt{2}}\mathrm{Re}(\mathcal{S}_{-2}^2+\mathcal{S}_{2}^2)$. This part is zero for a light field linearly polarized with $\beta=0$, as at the center of the trap, but is not quite zero at other positions across the trap volume. The component can split the energies of the $m_N=\pm1$ states but its size is comparatively small---note the $\times 40$ scaling---and so has no significant effect on the trapping potential. This is confirmed by the nearly identical trapping frequencies calculated for the two states at the beginning of this section.

\section{Cooling recipe and conclusions}
\label{sec:conclusions}
We have shown how to apply sideband cooling techniques to laser-coolable $^2\Sigma$ molecules in optical tweezer traps. The cooling must proceed from the $N=1$ rotational level to avoid decays to other rotational levels, but the resulting state-dependent potentials introduce significant additional complexity. This complexity can be greatly reduced by applying a large magnetic field to decouple the nuclear and electron spins from the rotational angular momentum. Under these conditions, families of states can be found with three ground states coupled to a single excited state. For certain choices of laser polarization, two of the three ground states have equal ac Stark shifts, so the frequency of the Raman transition between them is independent of the motional state. The reduction to only three ground states also greatly reduces the number of photons scattered during the optical pumping step, thereby reducing the heating. We have derived a formula for the additional heating caused by the different ac Stark shift of the third state, and have calculated the branching ratio out of this family of states as a function of applied magnetic field. 

These considerations lead us to a recipe for Raman sideband cooling of laser-coolable molecules such as CaF. For trapping light with linear incident polarization, the recipe is a straightforward extension of the scheme proposed for the simple molecule in Sec.~\ref{sec:simple-molecule}: (i) Apply a weak magnetic field\footnote{For this initial optical pumping, the magnetic field should be large enough that Zeeman splittings are large relative to any off-diagonal matrix elements of the tensor Stark interaction, but not large enough to uncouple the angular momenta.} along the incident polarization vector of the trapping light. Apply $\pi$ and $\sigma^-$ polarized light relative to this axis to optically pump molecules into the stretched state $\ket{N=1, F=2, m_F=-2}=\ket{ m_N=-1,m_S=-1/2,m_I=-1/2}$. According to Fig.~\ref{fig:branching}, this state is part of the spin manifold with the smallest branching ratio to other spin manifolds. (ii) Increase the magnetic field to $\sim \SI{300}{\gauss}$ in order to decouple the angular momenta. (iii) Drive the Raman transition from $\ket{m_N=-1,m_S=-1/2,m_I=-1/2}\rightarrow\ket{m_N=1,m_S=-1/2,m_I=-1/2}$ on the red motional sideband. (iv) Reapply optical pumping light from (i) to return molecules to $\ket{m_N=-1,m_S=-1/2,m_I=-1/2}$. (v) Repeat (iii) and (iv) until the molecule reaches the motional ground state. 

The circularly polarized optical pumping beam must be orientated along the $\mathcal{B}$ field. The linearly polarized beam must be orthogonal to this and, as discussed in Sec.~\ref{sec:simple-molecule}, to minimize heating should also be perpendicular to the weakly confining optical axis of the trap. Using Eq.~(\ref{Eq:heating-op}) and the trap frequencies for CaF in a real tweezer calculated in Sec.~\ref{sec:real-tweezer}, the mean change in motional quantum number during the optical pumping step can be written

\begin{equation}
    \overline{\Delta n^{\rm op}} = \kappa + \rho \,n.
\end{equation}
Under these conditions, we find $(\kappa, \rho)$ equal to $(0.17, 0.03)$ parallel to the incident polarization vector, $(0.11, 0.02)$ perpendicular to both the incident polarization and the optical axis, and $(0.29, 0.02)$ parallel to the optical axis. These calculations show sideband cooling should be effective on the first red sideband for small $n$. Driving higher-order sidebands may be helpful for initial cooling of hot clouds, particularly along the optical axis where the Lamb-Dicke parameter is much larger. As noted in Sec.~\ref{sec:simple-molecule}, the cooling can also work for other polarization choices---the main requirement is that two of the three states have equal tensor shifts. For example, choosing a linear polarization at an angle close to $\beta_{\rm magic}$ can satisfy this requirement and also reduce the heating due to $\Theta_{\rm curv}$, which is useful at high $n$. We conclude that the heating due to state-dependent potentials is not a major obstacle for effective cooling.

Figure \ref{fig:branching} shows that, at 300~G, the branching ratio to other spin manifolds is about $4\times 10^{-3}$. Increasing the magnetic field reduces the branching ratio further, but only by a factor of 2 for realistic fields. Recalling that an average of 3 photons are scattered in the optical pumping step, we see that under these conditions, the cooling cycle can be applied 58 times before half the population is lost to a different spin manifold. For the tweezer parameters considered here, 58 cycles of sideband cooling on the first red sideband corresponds to an energy reduction of $\SI{600}{\micro\kelvin}$ in either of the radial directions, or $\SI{100}{\micro\kelvin}$ in the axial direction. It is straightforward to use higher-order sidebands, resulting in proportionally larger energy reductions. Considering that molecular samples with temperatures of 5~$\mu$K have already been demonstrated by free-space laser cooling~\cite{Cheuk2018,Caldwell2019}, we see that it is feasible to reach the ground-state with little loss. We also note that loss to other spin manifolds is not fatal, since the cooling process could be applied to each spin manifold. Moreover, the entire cooling process---beginning with the optical pumping step at low magnetic field---can be applied multiple times if that proved necessary. 

Our analysis here has focused on CaF, but the methods and conclusions also apply to other similar molecules amenable to laser cooling. The ability to cool these molecules to the ground-state of tweezer traps is a key advance that will open the door to molecules as processors of quantum information and simulators of many-body quantum systems.

\begin{acknowledgements}
We are grateful to Jeremy Hutson and Ed Hinds for helpful discussions. This work was supported by EPSRC under grants EP/M027716/1 and EP/P01058X/1.
\end{acknowledgements} 

\clearpage
\onecolumngrid
\appendix*

\section{Stark shift operator and its matrix elements}
\label{sec:polarisability-op}

\subsection{Operator}

Consider a diatomic molecule interacting with light which has electric field amplitude ${\cal E}_0$, angular frequency $\omega_{\rm L}$, and unit polarization vector $\epsilon$. The interaction Hamiltonian is $H'=-\vec{d}\cdot \vec{E}$, where $\vec{d}$ is the dipole moment operator of the molecule and $\vec{E}= \frac{1}{2} {\cal E}_0 (\epsilon e^{-i \omega_{\rm L} t} + \epsilon^{*} e^{i \omega_{\rm L} t})$ is the electric field. We suppose that all effects much larger than this molecule-light interaction are included in a zeroth-order Hamiltonian, $H_0$, while the molecule-light interaction and all effects of a similar (or smaller) size are treated by perturbation theory. In second-order perturbation theory, the energy shift of a non-degenerate level $i$ is

\begin{equation}
\Delta W_i=-\frac{1}{4} \mathcal{E}_0^2 \sum _{j \ne i} \left(\frac{\bra{i} \vec{d}\cdot\epsilon ^*\ket{j} \bra{j}\vec{d}\cdot\epsilon \ket{i} }{\hbar  \left(\omega _{j i}-\omega_{\rm L} \right)}+\frac{\bra{i}\vec{d}\cdot\epsilon \ket{j}\bra{j}\vec{d}\cdot\epsilon ^*\ket{i} }{\hbar  \left(\omega _{j i}+\omega_{\rm L} \right)}\right)
\label{Eq:DW1}
\end{equation}
Here, $\omega_{j i}$ is the transition angular frequency between states $j$ and $i$ and the sum is over all states of the molecule. More generally, we may wish to know the energy shift of levels that are degenerate in the absence of the light, or handle cases where the ac Stark shifts are comparable to other level shifts and splittings, such as those arising from the hyperfine or Zeeman interactions. What is needed is an effective operator, which we will call $H_{\rm S}$, that describes the effect of the light within a small subspace of levels, for example, a single rotational state. The matrix elements of the effective operator between states $\ket{i}$ and $\ket{i'}$ within the subspace are the generalization of Eq.~(\ref{Eq:DW1}):

\begin{equation}
\bra{i} H_{\rm S} \ket{i'} = -\frac{1}{4} \mathcal{E}_0^2 \sum _{j} \left(\frac{\bra{i} \vec{d}\cdot\epsilon ^*\ket{j} \bra{j}\vec{d}\cdot\epsilon \ket{i'} }{\hbar  \left(\omega _{j i}-\omega_{\rm L} \right)}+\frac{\bra{i}\vec{d}\cdot\epsilon \ket{j}\bra{j}\vec{d}\cdot\epsilon ^*\ket{i'} }{\hbar  \left(\omega _{j i}+\omega_{\rm L} \right)}\right)
\label{Eq:Heff1}
\end{equation}
where the sum is over all states of $H_0$ that lie outside the subspace \cite{*[{A derivation of this result can be found in, for example, }] [{. See equation (26) of Complement B$_{\rm I}$. Note that the zeroth-order energies include the energies of the photons in the light field, and that the perturbation couples states that change the photon number by $\pm 1$.}] Cohen-Tannoudji1992}. In spherical coordinates, this expression is
\begin{equation}
\bra{i} H_{\rm S} \ket{i'}=-\frac{\mathcal{E}_0^2}{4}\sum _{p,q}(-1)^{p+q}\left(\bra{i} d_p {\mathcal R^{-}}d_q\ket{i'}(\epsilon^*)_{-p} \epsilon_{-q}+ \bra{i}d_q {\mathcal R^{+}}d_p\ket{i'}  \epsilon _{-q}(\epsilon^*)_{-p}\right)
\label{Eq:DW2}
\end{equation}
where we have defined the operator
\begin{equation}
{\mathcal R^{\pm}} =\sum_{j} \frac{1}{\hbar (\omega_{j i}\pm \omega_{\rm L})}|j\rangle\langle j|.
\label{EqApp:Resolvent}
\end{equation}
Provided $H_0$ does not include external fields, ${\mathcal R^{\pm}}$ is invariant under rotations. 

The formula for building a spherical tensor of rank $k_{12}$ from the product of two other spherical tensors of ranks $k_1$ and $k_2$ is
\begin{equation}
T_{p_{12}}^{k_{12}}(A,B)=\sum _{p_1} (-1)^{k_1+k_2-p_{12}}\sqrt{2 k_{12}+1} \, T_{p_1}^{k_1}(A)\, T_{p_{12}-p_1}^{k_2}(B) \left(
\begin{array}{ccc}
k_1 & k_2 & k_{12} \\
p_1 & p_{12}-p_1 & -p_{12} \\
\end{array}
\right).
\end{equation}
Applying this to the tensor product of two vectors, $\vec{u}$ and $\vec{v}$ gives
\begin{equation}
T_{P}^{K}(u,v)=(-1)^{P}\sqrt{2 K+1}\sum _{p}T_{p}^1(u)T_{P-p}^1(v)\left(
\begin{array}{ccc}
1 & 1 & K \\
p & P-p & -P \\
\end{array}
\right).
\end{equation}
The inverse relation gives us the expansion of the product $u_p v_q$ as
\begin{equation}
u_p v_q=\sum _{K=0}^2 \sum _{P=-K}^K (-1)^P\sqrt{2 K+1}\left(
\begin{array}{ccc}
1 & 1 & K \\
p & q & -P \\
\end{array}
\right)T_P^K(u,v).
\label{Eq:upvq}
\end{equation}

Equation (\ref{Eq:DW2}) contains two products of this form, one relating to the transition dipole moments of the molecule, and the other to the polarization of the light. Expanding each using Eq.~(\ref{Eq:upvq}), then evaluating the sums over $p$ and $q$, we find that
\begin{equation}
\sum _{p,q}(-1)^{p+q} \left(d_p {\mathcal R^{\pm}} d_q\right) \left(\epsilon ^*{}_{-p} \epsilon _{-q}\right)=\sum _{K=0}^{2}\sum_{P=-K} ^{K}(-1)^PT_P^K(d,{\mathcal R^{\pm}}d)T_{-P}^K\left(\epsilon ,\epsilon ^*\right).
\end{equation}
Note that the transformation $d_p {\mathcal R^{\pm}} d_q \rightarrow d_q {\mathcal R^{\pm}} d_p$ on the left-hand side of this equation multiplies the terms in the sum over $K$ on the right-hand side by $(-1)^{K}$. Applying these results to Eq.~(\ref{Eq:DW2}), we find that the effective Stark shift operator is
\begin{equation}
H_{\rm S}=-\frac{\mathcal{E}_0^2}{4}\sum _{K=0}^2 \sum _{P=-K}^K  (-1)^P {\mathcal A}^{K}_P \mathcal{P}_{-P}^K
\label{EqApp:HStark}
\end{equation}
where we have introduced the polarizability operators
\begin{subequations}
\begin{align}
{\mathcal A}^{K}_P &= ({\mathcal A^-})^{K}_P + (-1)^K ({\mathcal A^+})^{K}_P, \\
({\mathcal A^{\pm}})^{K}_P &= \frac{1}{z_K}T^{K}_P(d,{\mathcal R^{\pm}} d),
\end{align}
\label{EqApp:AKP}
\end{subequations}
and the polarization tensors
\begin{equation}
{\mathcal P}^{K}_P =z_K T^{K}_P(\epsilon,\epsilon^*).
\label{EqApp:PKP}
\end{equation}
The $z_K$ are numerical factors that can be chosen arbitrarily. We choose $z_0=-\sqrt{3}$ so that $\mathcal{P}_0^0 = \epsilon\cdot\epsilon^* = 1$, $z_1=-\sqrt{2}$ so that $\mathcal{P}^1 = -i(\epsilon \times \epsilon^*)$, and $z_2=\sqrt{3/2}$, so that $\mathcal{P}^2_0 = 1$ for light that is linearly polarized along $z$.

The components of ${\mathcal A^{\pm}}$ are
\begin{subequations}
	\label{Eq:alphaTensor}
\begin{align}
({\mathcal A^{\pm}})_0^0&=\frac{1}{3}\left(-d_1{\mathcal R^{\pm}}d_{-1}+d_0{\mathcal R^{\pm}}d_0-d_{-1}{\mathcal R^{\pm}}d_1\right), \\
({\mathcal A^{\pm}})_0^1&=\frac{1}{2}\left(-d_1{\mathcal R^{\pm}}d_{-1}+d_{-1}{\mathcal R^{\pm}}d_1\right), \\
({\mathcal A^{\pm}})_{\pm 1}^1&=\pm \frac{1}{2}\left(d_0{\mathcal R^{\pm}}d_{\pm 1}-d_{\pm 1}{\mathcal R^{\pm}}d_0\right),\label{EqApp:A11}\\
({\mathcal A^{\pm}})_0^2&=\frac{1}{3}\left(d_1{\mathcal R^{\pm}}d_{-1}+2d_0{\mathcal R^{\pm}}d_0+d_{-1}{\mathcal R^{\pm}}d_1\right),\\
({\mathcal A^{\pm}})_{\pm 1}^2&=\frac{1}{\sqrt{3}}\left(d_0{\mathcal R^{\pm}}d_{\pm 1}+d_{\pm 1}{\mathcal R^{\pm}}d_0\right),\\
({\mathcal A^{\pm}})_{\pm 2}^2&=\sqrt{\frac{2}{3}}d_{\pm 1}{\mathcal R^{\pm}}d_{\pm 1}.
\end{align}
\end{subequations}
The components of ${\cal P}$ are
\begin{subequations}
	\label{Eq:PTensor}
	\begin{align}
	\mathcal{P}_0^0&=\epsilon _{-1} \epsilon _{-1}{}^*+\epsilon _0 \epsilon _0{}^*+\epsilon _1 \epsilon _1{}^* = \epsilon\cdot\epsilon^*=1,\label{SEq:PTensor00}\\
	\mathcal{P}_0^1&=\epsilon _1 \epsilon _1{}^*-\epsilon _{-1} \epsilon _{-1}{}^*,\\
	\mathcal{P}_{\pm 1}^1&=\mp \left(\epsilon _0 \epsilon _{\mp 1}{}^*+\epsilon _0{}^* \epsilon _{\pm 1}\right),\\
	\mathcal{P}_0^2&=-\frac{1}{2}\left(\epsilon _{-1} \epsilon _{-1}{}^*-2 \epsilon _0 \epsilon _0{}^*+\epsilon _1 \epsilon _1{}^*\right)=-\frac{1}{2} \left(1-3 \epsilon _0 \epsilon _0{}^*\right),\\
	\mathcal{P}_{\pm 1}^2&=\frac{\sqrt{3}}{2}\left(-\epsilon _0\epsilon _{\mp 1}{}^*+\epsilon _0{}^* \epsilon _{\pm 1}\right),\\
	\mathcal{P}_{\pm 2}^2&=-\sqrt{\frac{3}{2}}\epsilon _{\mp 1}{}^* \epsilon _{\pm 1}
	\end{align}
\end{subequations}
Here, $\epsilon_p{}^{*}$ means $(\epsilon_p)^*$, and we have used the relation $(\epsilon^*)_q = (-1)^q (\epsilon_{-q})^*$.

\subsection{Matrix elements for \texorpdfstring{$^{1}\Sigma$}{¹Σ} states}\label{app:1S-states}

We consider a ground-state molecule with no orbital angular momentum, no electronic spin and no nuclear spin. In this simple case, the basis states are $|\Lambda, N, m_N\rangle$ with $\Lambda=0$. Here, the quantum numbers are the projection of the orbital angular momentum onto the internuclear axis ($\Lambda$), the rotational angular momentum ($N$), and its projection onto the $z$-axis ($m_N$). The matrix elements of the polarizability tensor are
\begin{equation}
\left\langle \Lambda,N',m_N'\left|{\mathcal A}_P^K\right|\Lambda ,N,m_N\right\rangle =(-1)^{N'-m_N'} \left(
\begin{array}{ccc}
N' & K & N \\
-m_N' & P & m_N \\
\end{array}
\right) \left\langle \Lambda,N'|| {\mathcal A}^K|| \Lambda ,N\right\rangle
\end{equation}
To evaluate the reduced matrix element, we rotate into the frame of the molecule using
\begin{equation}
{\mathcal A}_P^K=\sum _Q  \left(\mathcal{D}_{P Q}^K\right){}^* {\mathcal A}_Q^K.
\end{equation}
Here, the index $P$ is used for lab-frame components, and the index $Q$ for molecule-frame components, and $\mathcal{D}^K$ is the rotation operator of rank $K$ that transforms between them. This gives
\begin{align*}
&\left\langle \Lambda,N',m_N'\left|{\mathcal A}_P^K\right|\Lambda ,N,m_N\right\rangle =(-1)^{N'-m_N'}\left(
\begin{array}{ccc}
N' & K & N \\
-m_N' & P & m_N \\
\end{array}
\right)\sum _Q \left\langle \Lambda \left|{\mathcal A}_Q^K\right|\Lambda \right\rangle \left\langle \Lambda ,N'\left\|\left(\mathcal{D}_{.Q}{}^K\right){}^*\right\|\Lambda ,N\right\rangle\\
&=(-1)^{N'-m_N'} \left(
\begin{array}{ccc}
N' & K & N \\
-m_N' & P & m_N \\
\end{array}
\right) \sqrt{(2 N+1) \left(2 N'+1\right)} (-1)^{N'-\Lambda } \left(
\begin{array}{ccc}
N' & K & N \\
-\Lambda & 0 & \Lambda  \\
\end{array}
\right) \left\langle \Lambda \left|{\mathcal A}_{Q=0}^K\right|\Lambda \right\rangle.
\end{align*}
In the first line, the dot in the subscript of the rotation operator indicates that the matrix element is reduced relative to the index $P$. In the last step, we've set $Q=0$ since this is the only-non-zero term in the sum over $Q$. Let us define the molecule-frame parallel and perpendicular polarizability components:
\begin{subequations}
	\label{Eq:alphaparperp}

\begin{align}
\alpha _{\|}&=\sum _j \left(\frac{1}{\hbar  \left(\omega _{ji}+\omega_{\rm L} \right)}+\frac{1}{\hbar  \left(\omega _{ji}-\omega_{\rm L} \right)}\right)|\left\langle X\left|d_0\right|j,\Sigma \right\rangle |^2\\
\alpha _{\perp}&=\sum _k \left(\frac{1}{\hbar  \left(\omega _{ki}+\omega_{\rm L} \right)}+\frac{1}{\hbar  \left(\omega _{ki}-\omega_{\rm L} \right)}\right)|\left\langle X \left|d_1\right|k,\Pi \right\rangle |^2
\end{align}
\end{subequations}
Here, $X$ labels the $^1\Sigma$ ground state of interest, the index $j$ labels the set of excited $\Sigma$ states, $k$ labels the set of excited $\Pi$ states, and the dipole operators are acting in the molecule frame. We note that $\left|\left\langle X \left|d_{-1}\right|k,\Pi \right\rangle \right|^2=\left|\left\langle X \left|d_1\right|k,\Pi \right\rangle \right|^2$ because a $\Pi$ state is an equal superposition of $\Lambda=\pm 1$. We introduce the molecular parameters $\alpha_K=\bra{\Lambda=0}{\mathcal A} _{Q=0}^K\ket{\Lambda=0} $, which we can think of as the scalar, vector and tensor polarizabilities in the molecular frame. Using the definitions for the components of ${\mathcal A}$ given by Eq.~(\ref{Eq:alphaTensor}), the definitions of $\alpha_{\|}$, $\alpha_{\perp}$ and $\alpha_K$, and the relation $\bra{i}d_q\ket{j} =(-1)^q \bra{j}d_{-q}\ket{i}$, we find the complete expression for the matrix elements (for $\Lambda=0, S=0$):
\begin{equation}
    \left\langle \Lambda,N',m_N'\left|{\mathcal A}_P^K\right|\Lambda ,N,m_N\right\rangle = (-1)^{m_N'} \left(
\begin{array}{ccc}
N' & K & N \\
-m_N' & P & m_N \\
\end{array}
\right) \sqrt{(2 N+1) \left(2 N'+1\right)} \left(
\begin{array}{ccc}
N' & K & N \\
-\Lambda & 0 & \Lambda  \\
\end{array}
\right) \alpha_K,
\end{equation}
where
\begin{subequations}
\begin{align}
\alpha_0 &=\frac{1}{3}\left(\alpha _{\|}+2\alpha _{\perp}\right),\\
\alpha_1 &=0,\\
\alpha_2 &=\frac{2}{3}\left(\alpha _{\|}-\alpha _{\perp}\right).
\end{align}
\end{subequations}

\subsection{Matrix elements for \texorpdfstring{$^{2}\Sigma$}{²Σ} states}\label{app:2S-states}

Now we consider a more complicated case where the basis states are $|\Lambda, N, S, J, I, F, m_F\rangle$. In this order, the quantum numbers are the projection of the orbital angular momentum onto the internuclear axis, the rotational angular momentum, the total electronic spin, the total electronic angular momentum, the nuclear spin, the total angular momentum, and the projection of the total angular momentum onto the $z$-axis. Later, we shall also introduce the quantum numbers $\Sigma$ and $\Omega$, which are the projections of $\vec{S}$ and $\vec{J}$ onto the internuclear axis. Using the Wigner-Eckart theorem, and the fact that the operator acts in the space of the electronic coordinates, we have
\begin{align*}
&\left\langle \Lambda ,N',S,J',I,F',m_F'\left|{\mathcal A}_P^K\right|\Lambda ,N,S,J,I,F,m_F\right\rangle \\
&= (-1)^{F'-m_F'} (-1)^{F+J'+K+I}\sqrt{(2 F+1) \left(2 F'+1\right)}  \left(
\begin{array}{ccc}
F' & K & F \\
-m_F' & P & m_F \\
\end{array}
\right) \left\{
\begin{array}{ccc}
J'& F'& I \\ F & J & K \\
\end{array}
\right\} \left\langle \Lambda ,N',S,J'\| {\mathcal A}^K\| \Lambda ,N,S,J\right\rangle.
\end{align*}
To help evaluate the remaining matrix element, we use the relation between Hund's case (b) and case (a) states:
\begin{equation}
|\Lambda ,N,S,J\rangle =\sum _{\Sigma =-S}^S \sqrt{2 N+1} (-1)^{N-S+\Omega } \left(
\begin{array}{ccc}
J & S & N \\
\Omega  & -\Sigma  & -\Lambda  \\
\end{array}
\right)|\Lambda ,S,\Sigma ,J,\Omega \rangle.
\end{equation}
At this point, we specialize to $^2\Sigma$ states, which have $\Lambda=0$ and $S=1/2$. We also use the fact that our operator is built up from dipole moment operators that cannot change $\Sigma$. So, our reduced matrix element can be expressed as
\begin{align}
&\left\langle \Lambda ,N',S,J'\| {\mathcal A}^K\| \Lambda ,N,S,J\right\rangle =(-1)^{N'+N}\sqrt{(2 N+1)(2N'+1)}\left(
\begin{array}{ccc}
J & 1/2 & N \\
-1/2 & 1/2 & 0 \\
\end{array}
\right) \left(
\begin{array}{ccc}
J' & 1/2 & N' \\
-1/2 & 1/2 & 0 \\
\end{array}
\right) \times\nonumber\\
&\left[\left\langle \Lambda =0,\Sigma =-1/2,J'\| {\mathcal A}^K\| \Lambda =0,\Sigma =-1/2,J\right\rangle +(-1)^{J'+J+N'+N-1}\left\langle \Lambda =0,\Sigma =1/2,J'\left\|{\mathcal A}^K\right\|\Lambda =0,\Sigma =1/2,J\right\rangle \right]
\label{EqApp:Horrid1}
\end{align}
This leaves us with reduced matrix elements of the general type $\left\langle \Lambda,\Sigma,J'\| {\mathcal A}^K\| \Lambda, \Sigma, J\right\rangle$. As before, we rotate into the molecule frame and factorize the result, to reach
\begin{align}
\left\langle \Lambda ,\Sigma ,J'\| {\mathcal A}^K\| \Lambda ,\Sigma  ,J\right\rangle =\sqrt{(2 J+1) \left(2 J'+1\right)} (-1)^{J'-\Sigma } \left(
\begin{array}{ccc}
J' & K & J \\
-\Sigma  & 0 & \Sigma  \\
\end{array}
\right) \left\langle \Lambda,\Sigma \left|{\mathcal A}_{Q=0}^K\right|\Lambda ,\Sigma  \right\rangle.
\label{EqApp:removeJ}
\end{align}
We can use this result in Eq.~(\ref{EqApp:Horrid1}), noting that the two matrix elements in the square brackets differ only in the sign of $\Sigma$, and that changing the sign of $\Sigma$ in the 3j symbol of Eq.(\ref{EqApp:removeJ}) introduces an extra phase factor of $(-1)^{J+J'+K}$. Thus we obtain
\begin{align}
    &\left\langle \Lambda ,N',S,J'\| {\mathcal A}^K\| \Lambda ,N,S,J\right\rangle =\nonumber\\
    &(-1)^{N'+N+2J'+J+K+1/2}\sqrt{(2 N+1)(2N'+1)(2 J+1)(2J'+1)}
    \left(
\begin{array}{ccc}
J & 1/2 & N \\
-1/2 & 1/2 & 0 \\
\end{array}
\right) \left(
\begin{array}{ccc}
J' & 1/2 & N' \\
-1/2 & 1/2 & 0 \\
\end{array}
\right) \left(
\begin{array}{ccc}
 J' & K & J \\
 -1/2 & 0 & 1/2 \\
\end{array}
\right) \nonumber\\
&  \times\left[\left\langle \Lambda =0,\Sigma =-1/2\left|{\mathcal A}_{Q=0}^K\right|\Lambda =0,\Sigma =-1/2\right\rangle + (-1)^{N'+N-K} \left\langle \Lambda =0,\Sigma =1/2\left|{\mathcal A}_{Q=0}^K\right|\Lambda =0,\Sigma =1/2\right\rangle  \right]
\end{align}

The molecule-frame matrix elements in square brackets can be evaluated using the definitions in Eq.~(\ref{Eq:alphaTensor}). They involve terms of the type $|\langle \Lambda=0,\Sigma=\pm 1/2|d_q|\Lambda'',\Sigma''\rangle|^2$, where the double-primes refer to excited electronic states, and the $d_q$ operate in the molecule frame. The terms involving $d_0$ connect to the excited $^2\Sigma$ states, while the terms involving $d_{\pm 1}$ connect to the $^2\Pi_{1/2}$ and $^2\Pi_{3/2}$ states. For example
\begin{equation}
    \bra{ \Lambda=0,\Sigma=-1/2} d_1{\mathcal R^{\pm}} d_{-1}\ket{\Lambda =0,\Sigma =-1/2} =-\sum _k \frac{1}{\hbar  \left(\omega _k\pm \omega_{\rm L} \right)}|\bra{ X,^2\Sigma_{-1/2}} d_1 \ket{k,^2\Pi_{-3/2}}|^2,
\end{equation}
where the sum is over all the $\Pi$ states. We can write similar expressions for $\Sigma = 1/2$ and for the other operators appearing in the ${\mathcal A}_{Q=0}^K$. Using these expressions, the term in the square brackets can be written in terms of the following quantities:
\begin{subequations}
	\begin{align}
	\alpha _{\|}&=\sum _j \left(\frac{1}{\hbar  \left(\omega _{j}+\omega_{\rm L} \right)}+\frac{1}{\hbar  \left(\omega _{j}-\omega_{\rm L} \right)}\right)|\left\langle X,^2\Sigma \left|d_0\right|j,^2\Sigma \right\rangle |^2,\\
	\alpha _{\perp,\Omega }&=\sum _k\left(\frac{1}{\hbar \left(\omega _{k,\Omega }+\omega_{\rm L} \right)}+\frac{1}{\hbar  \left(\omega _{k,\Omega }-\omega_{\rm L} \right)}\right)|\left\langle X,^2\Sigma \left|d_1\right|k,^2\Pi _{\Omega }\right\rangle |^2,\\
	\alpha_{\perp} &= \frac{1}{2}\left( \alpha _{\perp,\frac{1}{2}} + \alpha _{\perp,\frac{3}{2}}\right)\\
	\beta _{\perp,\Omega }&=\sum _k\left(\frac{1}{\hbar \left(\omega _{k,\Omega }+\omega_{\rm L} \right)}-\frac{1}{\hbar  \left(\omega _{k,\Omega }-\omega_{\rm L} \right)}\right)|\left\langle X,^2\Sigma \left|d_1\right|k,^2\Pi _{\Omega }\right\rangle |^2\nonumber\\
	&=\sum _k\left(\frac{1}{\hbar \left(\omega _{k,\Omega }+\omega_{\rm L} \right)}+\frac{1}{\hbar  \left(\omega _{k,\Omega }-\omega_{\rm L} \right)}\right)\frac{\omega_{\rm L} }{\omega _{k,\Omega }}|\left\langle X,^2\Sigma \left|d_1\right|k,^2\Pi _{\Omega }\right\rangle |^2.
	\end{align}
	\label{EqApp:alphaSumStates}
\end{subequations}

After some algebra, we end up with a complete expression for the matrix elements (for $\Lambda=0, S=1/2$):
\begin{align}
&\left\langle \Lambda ,N',S,J',I,F',m_F'\left|{\mathcal A}_P^K\right|\Lambda ,N,S,J,I,F,m_F\right\rangle =\left[(-1)^{N'+N}+1\right] (-1)^{F'-m_F'+F-J'+J+I+1/2}\sqrt{(2 F+1) \left(2 F'+1\right)} \times\nonumber\\
&\sqrt{(2 N+1) \left(2 N'+1\right)(2 J+1) \left(2 J'+1\right)} \left\{
\begin{array}{ccc}
J'& F'& I \\ F & J & K \\
\end{array}
\right\}  \left(
\begin{array}{ccc}
F' & K & F \\
-m_F' & P & m_F \\
\end{array}
\right)\left(
\begin{array}{ccc}
J & 1/2 & N \\
-1/2 & 1/2 & 0 \\
\end{array}
\right) \left(
\begin{array}{ccc}
J' & 1/2 & N' \\
-1/2 & 1/2 & 0 \\
\end{array}
\right) \times\nonumber\\
& \left(
\begin{array}{ccc}
J' & K & J \\
-1/2 & 0 & 1/2 \\
\end{array}
\right)\alpha_K,
\end{align}

where

\begin{subequations}
	\begin{align}
	\alpha _0&=\frac{1}{3}\left(\alpha _{\|}+\alpha _{\perp,\frac{1}{2}}+\alpha _{\perp,\frac{3}{2}}\right)=\frac{1}{3}\left(\alpha_{\|}+2\alpha_{\perp}\right)\label{EqApp:alpha0},\\
	\alpha _1&=\frac{1}{2}\left(\beta _{\perp,\frac{1}{2}}-\beta _{\perp,\frac{3}{2}}\right),\label{EqApp:alpha1}\\
	\alpha _2&=\frac{1}{3}\left(2\alpha _{\|}-\alpha _{\perp,\frac{1}{2}}-\alpha _{\perp,\frac{3}{2}}\right) = \frac{2}{3}\left(\alpha_{\|}-\alpha _{\perp}\right).\label{EqApp:alpha2}
	\end{align}
	\label{EqApp:alpha}
\end{subequations}

Note that in the particular case where only one $\Pi$ state contributes to the sum, $\beta_{\perp,\Omega }=\frac{\omega}{\omega_{m,\Omega }}\alpha_{\perp,\Omega}$.
In the special case where the detuning of the light is very large compared to the fine-structure interval of the excited state, $\alpha_{\perp,\frac{1}{2}} \approx \alpha_{\perp,\frac{3}{2}}$. When both of these special cases hold,

\begin{equation}
	\alpha _1\approx \frac{1}{2}\frac{\left(\omega _{3/2}-\omega _{1/2}\right) \omega }{\omega _{1/2} \omega _{3/2}}\alpha _{\perp}.
	\label{EqApp:alpha1Approx}
\end{equation}

\twocolumngrid
\bibliography{references}

\end{document}